\documentclass[onecolumn,runningheads,final,amssymb,amsmath,natbib]{svjour2}
\usepackage{epsfig}
\usepackage{graphicx}
\usepackage{epic}
\usepackage{eepic}
\usepackage{bbold}
\usepackage{mathptmx}
\journalname{Celestial Mechanics and Dynamical Astronomy}

\newcommand{\be}{\begin{equation}}
\newcommand{\ee}{\end{equation}}
\newcommand{\ba}{\begin{eqnarray}}
\newcommand{\ea}{\end{eqnarray}}
\newcommand{\RefEq}[1]{{(\ref{#1})}}

\begin{document}

\title{ Strands and braids in narrow planetary rings: A scattering system
  approach}
\titlerunning{Strands and braids in narrow rings}
\author{O. Merlo \and L. Benet}
\institute{O. Merlo \and L. Benet \at Instituto de Ciencias
  F\'{\i}sicas, Univeridad Nacional Aut\'onoma de M\'exico (UNAM),
  Apdo. Postal 48--3, 62251--Cuernavaca, Mor., M\'exico\\
  \email{merlo@fis.unam.mx, benet@fis.unam.mx}}

\date{\today}

\maketitle

\begin{abstract}
  We address the occurrence of narrow planetary rings and some of
  their structural properties, in particular when the rings are
  shepherded. We consider the problem as Hamiltonian {\it scattering}
  of a large number of non-interacting massless point particles in an
  effective potential. Using the existence of stable motion in
  scattering regions in this set up, we describe a mechanism in phase
  space for the occurrence of narrow rings and some consequences in
  their structure. We illustrate our approach with three examples. We
  find eccentric narrow rings displaying sharp edges, variable width
  and the appearance of distinct ring components (strands) which are
  spatially organized and entangled (braids). We discuss the relevance
  of our approach for narrow planetary rings.  
\end{abstract}


\section{Introduction}

The accidental discovery of the rings of Uranus in 1977 by
stellar-occultation measurements~\citep{Elliot77} led to a renewed
interest in planetary ring systems. The main reason was that the
Uranian rings turned to be extremely different from those of Saturn:
They are narrow, opaque, sharp-edged, inclined and
eccentric~\citep{Elliot84,Esposito02}. To quote some
figures~\citep{Murray99}, the widest ring of Uranus, the $\epsilon$
ring, is 20--96 km wide with a nominal semi-major axis at 51,149 km;
in comparison, the main rings of Saturn are a few thousand kilometers
wide. The eccentricity of the $\epsilon$ ring is $0.0079$; Saturn's F
ring has an eccentricity $0.0026$. The structural properties have
raised a number of questions, most of which remain
unanswered~\citep{Esposito02,Sicardy05}. For instance, an eccentric
inclined narrow ring like the $\epsilon$ ring is expected to spread in
rather short time scales, $t_{\rm max} = 6\times 10^{8}$ years, which
is ``considerably smaller than the age of the solar
system''~\citep{Esposito02}. This is due to interparticle collisions,
drag and differential precession. This suggests an efficient
confinement mechanism that maintains these structural properties of
the ring over longer time scales~\citep{Esposito02}.

To explain these structural features, new models were introduced where
the confinement was induced by nearby moons. Among these models we
mention in particular the shepherding model introduced by
\citet{Goldreich79}, where two moons around the ring were proposed to
bound it. Soon afterwards, the Pioneer and Voyager missions detected
the shepherd moons around Saturn's F ring, and later the shepherds of
the outermost $\epsilon$ ring of Uranus. The shepherding confinement
involves angular momentum transfer between the shepherd moons and the
ring particles, self-gravity and viscous damping due to interparticle
collisions~\citep{Borderies83}. While the full scenario for
shepherding has not been fully
understood~\citep{Esposito02,Sicardy05}, the presence of dissipation
seems to be essential. In addition, the formulation assumes the ring
boundary located at a lower-order resonance.

While the discoveries of the shepherd moons around the F Saturn ring
and the $\epsilon$ ring of Uranus represented some confirmation for
the shepherd theory, other questions still remain unanswered. For
instance, most Uranian rings have no associated shepherd moons around
them~\citep{Murray90}, nor the narrow eccentric rings of Saturn, which
among others would provide an explanation for their
sharp-edges~\citep{Murray99}. Thus, either the shepherds are there but
are too small to be detected, or ``some physics is yet to be
understood''~\citep{Sicardy05}. Other problems are related with the
effects of collisions, drag or keplerian differential velocity, which
define too short life-times for the narrow rings, and cast doubts on
the confinement mechanism which maintains the eccentricity and sharp
edges of these rings. In addition, Saturn's F ring turned out to have a
very rich dynamical structure~\citep{Smith81,Smith82,Murray97}: besides the
non-zero eccentricity, it displays multiple components entangled in a
complicated way, known as strands and braids, showing further puzzling
features like kinks and clumps.

Numerical simulations have investigated a variety of physical
interactions, like the gravitational perturbations of shepherd moons
on circular and eccentric
orbits~\citep{ShowalterBurns1982,GiuliattiWinter00}, effects due to the
action of embedded moonlets~\citep{LissauerPeale1986}, and ring
interparticle collision
effects~\citep{Hanninen1993,LewisStewart2000}. The central questions
investigated have been the formation of structure (strands, braids,
clumps) and their short-term stability. While these studies are
extremely valuable and have led to interesting predictions, e.g. the
formation of channels and streamers~\citep{GiuliattiWinter00} which have
been recently observed~\citep{Murray05}, there is no self-consistent
approach for the confinement of narrow rings and their radial and
azimuthal structure. The F Saturn ring remains as the most fascinating
example of these possibilities.

The present paper addresses this point, namely, a self-consistent
scenario for the occurrence of narrow rings and the appearance of
structure. The main idea is to consider phase space regions where {\it
  scattering} dominates the dynamics in systems with some {\it
  intrinsic rotation}. Scattering provides a simple scenario where
confinement and escape are clearly distingushed. Existence of {\it
  stable} periodic orbits or tori in such scattering regions implies
effective (over long times) dynamically bounded motion where otherwise
there would be escaping trajectories. The intrinsic rotation, which is
associated to an external potential due to the shepherds or by the
full many--body problem, defines a mechanism for creating the ring:
the trapped region, if it exists, rotates around the central planet,
thus the confined particles form a ring. While at first sight it may
appear a strange idea to focus on scattering dynamics, close
approaches to shepherds illustrate this situation. In plain words,
what does not escape to infinity (due to scattering) builds a ring.
The bifurcation scenario yielding the stable periodic orbits or tori,
and other considerations supported by theorems of Hamiltonian systems
with many degrees of freedom, allow us to understand the narrowness
and the appearance of strands and braids. Within this approach, which
at this stage is mainly qualitative, we succeed in explaining the
confinement and also some of the questions described above, such as
the occurrence of narrow eccentric rings with sharp edges, apse
alignment, the appearance of different ring components (strands) and
their spatial entanglement (braids). The narrow rings we obtain for
the examples treated are stable although they may display dynamical
evolution.

To illustrate our approach we consider three examples. The first, a
planar billiard system (a disk rotating on a circular or Kepler
ellipse) with an arbitrary large number of non-interacting massless
particles, is very simple, yet the relevant steps can be followed to a
good extent analytically. So it is an interesting system that allows a
pedagogical insight on many properties of the scattering approach to
narrow rings. The second example in addition includes a central
attractive $1/r$ interaction; it helps to establish the connection to
the full gravitational problem, showing the genericity and robustness
of our approach. The third example is the circular restricted
three-body problem with an extremely small mass parameter. With these
examples we show the occurrence of rings, the appearance of radial and
azimuthal structures, and the genericity and robustness of the whole
approach.

The paper is organized as follows: in the next section we describe our
approach in rather general terms, avoiding formalities. In
Sec.~\ref{ex1} we illustrate our theory with the first example, a
scattering billiard system. The main steps in the construction of a
(narrow) ring are described in detail, discussing separately circular
and elliptic motion of the billiard disk. For the elliptic case we
present new results related to the appearance of structure within the
rings, in particular, strands and braids. In Sec.~\ref{ex2} we present
results for the rotating disk with the addition of a central
attractive $1/r$ interaction. This example permits us to understand
the connection to the gravitational problem, which we exemplify with
the circular restricted three-body problem in
Sec.~\ref{rtbp}. Finally, in Sec.~\ref{concl} we present conclusions
and some outlook.

\section{Hamiltonian scattering approach to rings of non-interacting 
  particles}
\label{theory}

We begin by considering the planar circular restricted three-body
problem (RTBP) of point particles with a very small mass
parameter~\citep{Szebehely67}; more complicated and realistic cases
will be considered below. The larger mass (primary) is thus associated
to the central planet, the smaller one (secondary) to a confining
moon, and the massless particle to a particle of the ring, and
therefore it does not influence the motion of the other two. The
motion of the massive bodies is circular about their common center of
mass. In an inertial frame with the origin at the central planet, the
Hamiltonian describing the motion of the massless particle takes the
general form
\be
\label{eq1}
H = \frac{1}{2}(P_X^2+P_Y^2) + V_0(|\vec{X}|) 
  + V_1(|\vec{X}-\vec{X}_d(\phi)|).
\ee
Here, $\vec{X}=(X,Y)$ denotes the position of the massless particle,
$P_X$ and $P_Y$ are the canonically conjugate momenta,
$\vec{X}_d(\phi)$ denotes the position of the secondary, and
$V_0(|\vec{X}|)$ and $V_1(|\vec{X}-\vec{X}_d(\phi)|)$ are the
interaction potentials due to the primary and secondary,
respectively. The Hamiltonian is explicitly time-dependent through the
dependence of $V_1$ upon $\phi = \omega t$, with $\phi$ the angular
position of the secondary defined with respect to the $X$-axis and
$\omega$ its constant angular velocity. By canonically changing to a
rotating frame (with constant angular velocity $\omega$), the new
Hamiltonian becomes time-independent and is thus a constant of motion
(cf. Eq.~\RefEq{eq2} below). This is the well-known Jacobi
constant~\citep{Szebehely67}.

We observe that the Hamiltonian~\RefEq{eq1} has a quite generic
form. It consists of an integrable part $H_0 = H - V_1(X,Y,\phi)$ and
a periodic time-dependent perturbation $V_1(X,Y,\phi)$. Following the
standard terminology of scattering theory~\citep{Newton}, we shall
denote by $H_0$ the {\it free} part of the interaction, which is
completely integrable. In the case of the RTBP, the potential $V_0$ is
proportional to the mass of the central planet, while $V_1$, the {\it
  disturbing} potential, is proportional to the mass parameter
$\mu$. For the known shepherd moons we have $\mu\sim
10^{-8}-10^{-10}$~\citep{Uralskaya03}. It is thus legitimate to
consider the motion as small corrections about a Kepler solutions
($V_0$), i.e., in a series expansion. In this case, we say that $V_0$
defines the geodesics of the free interaction $H_0$ (the conic
solutions for the Kepler problem), while $V_1$ acts as a small
perturbation.

A large variety of periodic orbits (in the rotating frame) exist for
the circular RTBP with small $\mu$~\citep{Henon97}. Considering the
measured eccentricities of the planetary narrow rings ($\le 0.01$,
cf.~\citealt{Murray99}), we are thus interested in those which have
a very small eccentricity, and therefore are close to the circular or
elliptic orbits of the two-body Kepler problem. In addition, these
orbits must occur in regions in phase space where scattering motion
and thus the possibility to escape to infinity dominates the dynamics,
i.e., no zero-velocity curves may exist~\citep{BST98}. That is, we
focus on phase space regions with no ``potential barriers'' that
confine the motion of the massless particle. It can be shown that such
periodic orbits indeed exist, and some are linearly stable in certain
intervals of the Jacobi constant~\citep{Henon97,BST98}. Moreover, these
stable periodic orbits appear through saddle--center bifurcations by
changing the value of the Jacobi integral (e.g., by decreasing it): a
pair of new periodic orbits is created at certain value of the Jacobi
constant, one stable and one unstable (see Fig.~\ref{fig3}a below). By
slowly changing the value of the Jacobi integral, the local horseshoe
changes with respect to the parameters that characterize
it~\citep{BST98}. Eventually, further reducing the Jacobi integral
induces a period-doubling bifurcation cascade on the stable periodic
orbit, which finally turns it into an unstable one
(cf. Fig.~\ref{fig3}b); further reduction yields locally a hyperbolic
horseshoe. The key property here is that, in the intervals of the
Jacobi integral where there is one stable periodic orbit, the
manifolds of the unstable partner define barriers which {\it confine
  dynamically} the motion of the massless particle
(Fig.~\ref{fig3}a). Such dynamical confinement actually takes place
all around the central planet due to the intrinsic rotation induced by
the motion of the shepherd. Hence, there is a region in phase space
where the solutions remain close to the stable periodic orbit, not
escaping, irrespective of whether the actual motion is periodic,
quasi-periodic or even chaotic. This behavior is generic for
autonomous two degrees-of-freedom scattering Hamiltonian systems of
the type considered; systems with more degrees of freedom are
discussed below.

We turn now to the occurrence of rings~\citep{Benet00}. Consider an
arbitrarily large number (ensemble) of massless particles whose
interaction is given by~\RefEq{eq1}, but which do not interact between
themselves. While the latter assumption is very restrictive, it is a
first step in the theory. It essentially allows us to separate the
many-body problem into a collection of independent one-particle
problems, which is the first term in the series expansion of the full
many-body Hamiltonian in terms of the mass of the ring particles. The
initial conditions of such non-interacting particles are chosen close
to the phase space location of the stable periodic solution described
above. Yet, they differ slightly, in particular in the value of the
Jacobi integral. We emphasize that we are not interested in collision
orbits with a shepherd but in periodic orbits very close to it.

We restrict now to one interval of Jacobi integrals where the whole
bifurcation scenario takes place (from the creation of the pair of
periodic orbits, until both define a hyperbolic structure). For a
fixed value of the Jacobi integral, all particles whose initial
conditions are outside the phase space region enclosed by the
manifolds of the unstable periodic orbit rapidly escape to infinity
along a scattering trajectory. In contrast, all those within the
manifolds are {\it dynamically} confined, moving around the central
planet. The distinction among these two cases is sharp. Thus, letting
the (non-interacting) system evolve from an initial time $t=0$ a large
number of particles escape, while the rest move close to the reference
stable solution. These statements hold for all values of the Jacobi
constant within the considered interval. The ring is then obtained, in
the $X$--$Y$ space for a given time, from the position of all the ring
particles of the ensemble that remain close to the reference periodic
orbit, i.e., that are dynamically confined (Figs.~\ref{fig4},
\ref{fig7} and~\ref{fig8}). Note that this argument does not require a
priori any resonance condition.

Some important properties of the ring are the following. First, the
ring displays sharp edges. This follows from considering only
scattering regions in phase space. Indeed, the distinction between
dynamically trapped particles and those with unbounded motion is
rather clear typically after a few periods of revolution of the
secondary mass. Yet, for very small values of the mass parameter
certain subtleties arise, as we shall discuss in sections~\ref{ex2}
and~\ref{rtbp}, due to the overwhelming attraction of the central
planet. Second, the rings are in general eccentric. This can be
understood by observing that the motion of each particle of the ring
is close to the originating stable periodic orbit, which in turn is
close to a keplerian ellipse of small eccentricity in the case of the
RTBP, as we have chosen it. In the rotating frame, the latter has a
well-defined periapse and apoapse. In fact, the ring resembles the
shape of the periodic orbit pictured in the rotating frame. Notice
that this is a collective property reflected in the $X$--$Y$ space, in
the sense that it is defined by the ensemble of ring particles which
are trapped. Put differently, apse alignment follows from the
proximity and resemblance of the trapped trajectories to the
organizing reference orbit (stable periodic orbit). Third, the
narrowness of the ring can be understood from the fact that the region
in phase space (for a specific Jacobi integral) corresponding to
dynamically bounded motion is typically quite small~\citep{BST98}, as
well as the interval of values of the Jacobi integral where the
reference periodic orbit is stable. This is usually so also because of
the presence of other moonlets. We emphasize that, despite the
qualitative nature of these results, these properties are observed in
real narrow planetary rings, and some of them remain as open
questions~\citep{Esposito02,Sicardy05}.

Above, we have used the fact that the Hamiltonian~\RefEq{eq1} is
autonomous with two degrees of freedom. This permits us to compute
directly the stability properties of the periodic orbits from the
trace and the determinant of the linearized
dynamics~\citep{LinStab}. Moreover, in this case the manifolds of the
unstable periodic orbit indeed define a region of bounded motion if
the partner is stable. One consequence is that, for the interval of
Jacobi constant of interest, the probability of being trapped is small
but {\it strictly} non--zero; this follows from the KAM
structure~\citep{Arnold88} for a two degree-of-freedom system
(Fig.~\ref{fig3}a). Furthermore, the scenario is structurally stable,
i.e., there are no qualitative changes under small generic
perturbations: the organizing centers (periodic orbits) will only be
shifted in phase space. Therefore, oblateness of the planet or
other effects can also be included in this scenario. The whole
approach, so far, is robust.

The question now is whether this approach can be extended to systems
with more degrees of freedom, for instance, when other shepherd moons
are also taken into account, as in the case of Saturn's F ring or the
$\epsilon$ ring of Uranus. For more degrees of freedom new phenomena
appear, such as Arnold diffusion~\citep{Arnold64}. Furthermore, the
larger dimensionality of phase space allows for other possibilities as
regards the stability of the periodic orbits~\citep{Skokos01}. These
properties seem to be a real constraint for extending our approach to
more degrees of freedom. Yet, for many degrees of freedom, the same
line of reasoning can be followed with some minor changes: instead of
presenting the theory on the stable periodic orbits, we do it on the
stable quasi-periodic tori. According to a theorem of
\citeauthor{Jorba97a} \citeyearpar{Jorba97a,Jorba97b}, around
lower-dimensional normally elliptic tori there is a region of {\it
  effective stability}. Moreover, if the corresponding unstable
solution is of center $\times$ center $\times\cdots\times$ center
$\times$ saddle type, recent results by Wiggins and collaborators
assert that the scattering dynamics can be understood similarly to the
invariant construction of the autonomous two-degree-of-freedom
system~\citep{Wiggins01,Uzer02,Waalkens04}. These results suggest that,
at least locally and in the sense of effective stability, i.e. for
long enough times but perhaps not infinitely long, the underlying
phase space structure is essentially the same as for two degrees of
freedom. In this case, the construction of the ring can be carried
out, even though we cannot assure infinitely long confinement. A
simple case where these statements can be checked is the planar
elliptic RTBP for small but non-zero eccentricity. In this case the
explicit quasi-periodic time-dependence of the potential $V_1$ cannot
be removed, leading to a phase space of larger dimensionality. As we
shall illustrate below, a quasi-periodic perturbation $V_1$ has
important consequences for the ring structure: the ring turns out to
have many components (see Fig.~\ref{fig5}a) which are entangled in a
complicate way (cf. Figs.~\ref{fig5}b--d).

We finish this section by emphasizing that, despite the fact that we
have considered the case of a shepherd moon for concreteness, the
intrinsic rotation defined by the circular orbit above, or any other
rotating quasi-periodic motion, may be associated with other moons or
with the effect of the whole many-body interacting problem. If this
intrinsic rotation allows for the appearance of trapped motion
embedded in scattering regions, which are somewhat localized, then
narrow rings will occur. This is therefore a possible explanation, not
requiring undiscovered shepherd moons, for the non-shepherded narrow
rings of Uranus for instance.

\section{Example 1: A scattering billiard system}
\label{ex1}

We illustrate our theory first with a planar billiard system. While
this system is a simple toy model, we avoid the numerical
complications of dealing with $1/r$ potentials where the relevant
masses differ significantly (many orders of magnitude). Yet, the
connection to the gravitational case can be established rigorously
(Sec.~\ref{ex2}). This toy model, as we shall show, has the great
advantage that it provides insight into the scattering properties on
which our approach relies, since the fact that there is no global
confining potential enhances the possibility of escape.

\begin{figure}
  \centerline{\includegraphics[angle=90,width=6cm]{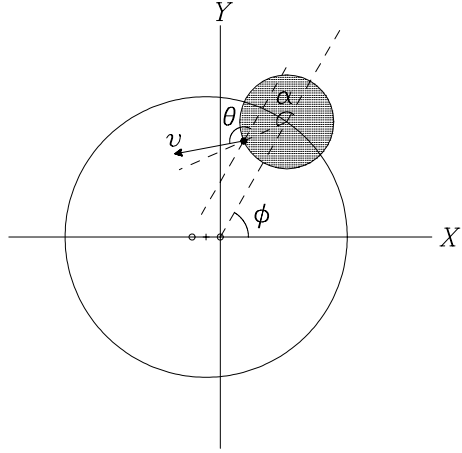}}
  \caption{ The scattering billiard: $\phi$ is the position of the
    center of the disk, $\alpha$ denotes the position of the collision
    point on the disk, $v$ is the magnitude of the outgoing velocity
    and $\theta$ defines its direction. The center of the disk moves
    on a Kepler ellipse, whose foci are shown on the $X$-axis as
    open circles ($\circ$), and its center by a cross~($+$).}
\label{fig1}
\end{figure}

Consider the planar motion of a massless point particle bouncing off
one circular hard disk of radius $d$, which moves on a Kepler
orbit. The center of the disk $\vec{X}_d(\phi)$ describes a circular
or elliptic Kepler orbit with one focus at the origin
(Fig.~\ref{fig1}). We denote its radial position by $R(\phi)$, with
$\phi$ the angular position along the Kepler orbit measured from the
pericenter. Then,
$R(\phi)=a(1-\varepsilon^2)/(1+\varepsilon\cos(\phi))$ with $a=1$ is
the semi-major axis and $\varepsilon$ the eccentricity. The
Hamiltonian expressed in an inertial frame can thus be written as
Eq.~\RefEq{eq1}, with $V_0=0$, $V_1$ zero for $|\vec{X} -
\vec{X}_d(\phi)|^2 > d^2$ and infinite otherwise. We shall focus on
the case of small or zero eccentricity. For non-zero $\varepsilon$,
due to the explicit time dependence the system has two and a half
degrees of freedom and no constant of motion.

By performing a canonical transformation of~\RefEq{eq1} to appropriate
rotating--pulsating coordinates~\citep{Merlo04}, the disk is left at rest at
$\vec{x}_d=(\bar{R},0)$, where $\bar{R}=a$ is the major semi-axis of the
orbit of the center of the disk. The new Hamiltonian $J$ is then given by
\ba
\label{eq2}
J &=& \frac{1}{2}\frac{\bar{R}^2}{[R(\phi)]^2} (p_x^2+p_y^2)
  + V_0\left(\frac{R(\phi)}{\bar{R}}|\vec{x}|\right)  
  + V_1\left(\frac{R(\phi)}{\bar{R}}|\vec{x}-\vec{x}_d|\right) \nonumber\\
  & &  - \dot\phi(xp_y-yp_x) 
  - \dot\phi\frac{1}{R(\phi)}\frac{{\rm d}R(\phi)}{{\rm d}\phi}(xp_x+yp_y).
\ea
For $\varepsilon=0$ (circular orbit), $J$ is time independent and thus a
constant of motion (the Jacobi integral). We shall refer here to the numerical
value of $J$ in~\RefEq{eq2} as the Jacobi integral, even for non-zero
eccentricity where it is not a constant of motion. 

The dynamics of the billiard is straightforward. The particle moves
freely on a rectilinear trajectory of constant velocity (the potential
$V_0=0$) until it encounters the disk; if it never collides then it
escapes to infinity. When a collision takes place, the particle is
specularly reflected with respect to the local (moving) frame of the
disk. This defines the outgoing conditions at the disk, and the motion
is rectilinear and uniform again. The precise result of a collision
depends on the position on the disk where it occurs, the relative
velocities, and for non-vanishing $\varepsilon$, on the position
along the ellipse ($\phi$). Collisions taking place on the front of
the disk increase the (outgoing) kinetic energy of the particle, while
collisions on the back reduce it.

In the context of rings, we are interested in the particles which are
dynamically trapped, the ring particles. In the present example, this
can only be obtained through consecutive collisions with the disk.  It
is important to emphasize here that collisions with the disk do not
model or imply physical collisions with the actual shepherds; they
model interactions with the external potential that allows for
confinement.  A convenient description of this situation is given by
introducing the following quantities, defined at the collision point
(Fig.~\ref{fig1}). The angle $\phi$ defines the position of the center
of the disk with respect to the $X$ axis (inertial frame), and $v$ is
the magnitude of the velocity after the collision. The angle $\alpha$
denotes the angle formed by $\vec{X}_d(\phi)$ and the position vector
of the collision point referred to the center of the disk. Finally,
the angle $\theta$ is the outgoing direction of the velocity.

\subsection{Occurrence of rings: The circular case}
\label{sec3.1}

\begin{figure*}
  \centerline{\includegraphics[angle=90,width=11.5cm]{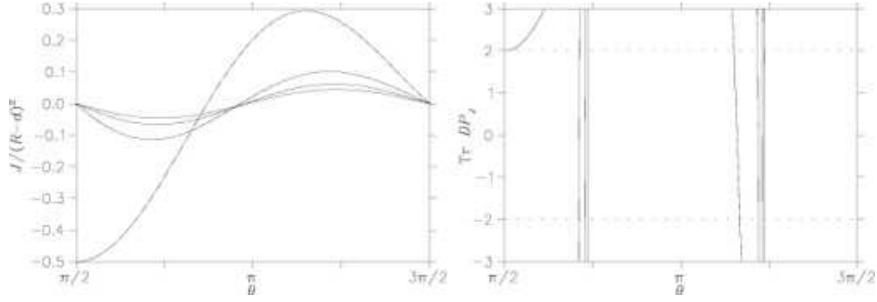}}
  \caption{ (a)~Radial (consecutive collision) periodic orbits of the
    circular rotating billiard for $n=0,1,2,3$, in terms of
    $J/(R-d)^2$. (b)~Stability of the radial periodic orbits: the
    orbits are stable if $| {\rm Tr\,} D{\cal P}_J| < 2$.}
\label{fig2}
\end{figure*}

For $\varepsilon=0$, Eq.~\RefEq{eq2} is an autonomous
two-degree-of-freedom Hamiltonian. The Hamiltonian flow defines a
map, e.g., at the collision point with the disk. This map is open, in
the sense that certain initial conditions may not have an associated
image, as due to escape. Trapped trajectories are associated with
consecutive collisions with the disk. In particular, simple
periodic orbits can be worked out analytically. The Jacobi integral is
given by $J = v^2/2 - v\left( R\sin\theta + d\sin(\theta-\alpha)
\right)$ ($\omega=1$). Due to the circular symmetry, all radial
collisions ($\alpha=\pi$) conserve the kinetic energy. It is easy to
calculate the condition that these orbits must fulfill to be fixed
points of the map~\citep{Meyer95},
\be
\label{eq3}
\frac{J_n}{ (R-d)^2 } 
   = \frac{2\cos^2\theta+\Delta\phi\sin(2\theta)}{(\Delta\phi)^2} 
   = \frac{2\cos^2\theta(1+\Delta\phi\tan\theta)}{(\Delta\phi)^2}.
\ee
Here, $n=0,1,\dots$ denotes the number of full turns completed by the
disk between consecutive collisions, and $\Delta\phi= (2n-1)\pi+2
\theta$ is the corresponding change in the angle $\phi$. In
Fig.~\ref{fig2}a we plot $J_n/(R-d)^2$ for some values of
$n$. However, the periodic orbits are a set of measure zero of the
initial conditions. Therefore, we must also consider the behavior of
nearby trajectories. To this end, we note that the characteristic
curves in Fig.~\ref{fig2}a display one maximum and one minimum for
each value of $n$. The occurrence of maxima and minima on these curves
points to the appearance of consecutive collision periodic orbits
through saddle--center bifurcations.

The stability properties of a time-independent two-degree-of-freedom
system can be obtained from the trace and the determinant of the
linearized Poincar\'e map at the fixed points~\citep{LinStab}. The
transformation $(\alpha_{k+1}, p_{k+1})={\cal P}_J(\alpha_k,p_k)$ with
$p_k=-d-R\cos\alpha_k-v\sin(\alpha_k-\theta_k)$ defines an iterative
canonical map. Here, $\alpha_k\in [0,2\pi]$ and
$p_k\in[-p_{\rm max},p_{\rm max}]$, where
$p_{\rm max}=(2J+R^2+d^2+2Rd\cos\alpha)^{1/2}$. Then, the information on
the stability is completely contained in the trace of the linearized
matrix $D{\cal P}_J$. One finds~\citep{Merlo04,Benet04}
\be
\label{eq4}
{\rm Tr\,} D{\cal P}_J = 2 +
\frac{(\Delta\phi)^2(1-\tan^2\theta)-4(1+\Delta\phi\tan\theta)}{d/R}.
\ee
Changes in the stability of the periodic orbits, i.e. bifurcations,
occur at ${\rm Tr\,} D{\cal P}_J =\pm 2$. Equating the r.h.s. of
Eq.~\RefEq{eq4} to $2$ turns out to be equivalent to the condition
${\rm d}J_n/{\rm d}\theta=0$~\citep{Benet00}, i.e., the condition
defining the position of the minima and maxima of $J_n$. The
corresponding result for $-2$ yields a condition related to
period-doubling bifurcations. Figure~\ref{fig2}b shows the behavior of
${\rm Tr\,} D{\cal P}_J$ in terms of $\theta$. Note that a simple
semi-analytical estimate of the width of the rings is obtained by
projecting the curves ${\rm Tr\,} D{\cal P}=\pm 2$ onto the $X$--$Y$
plane.

\begin{figure*}
  \centerline{
  \includegraphics[angle=270,width=5.5cm]{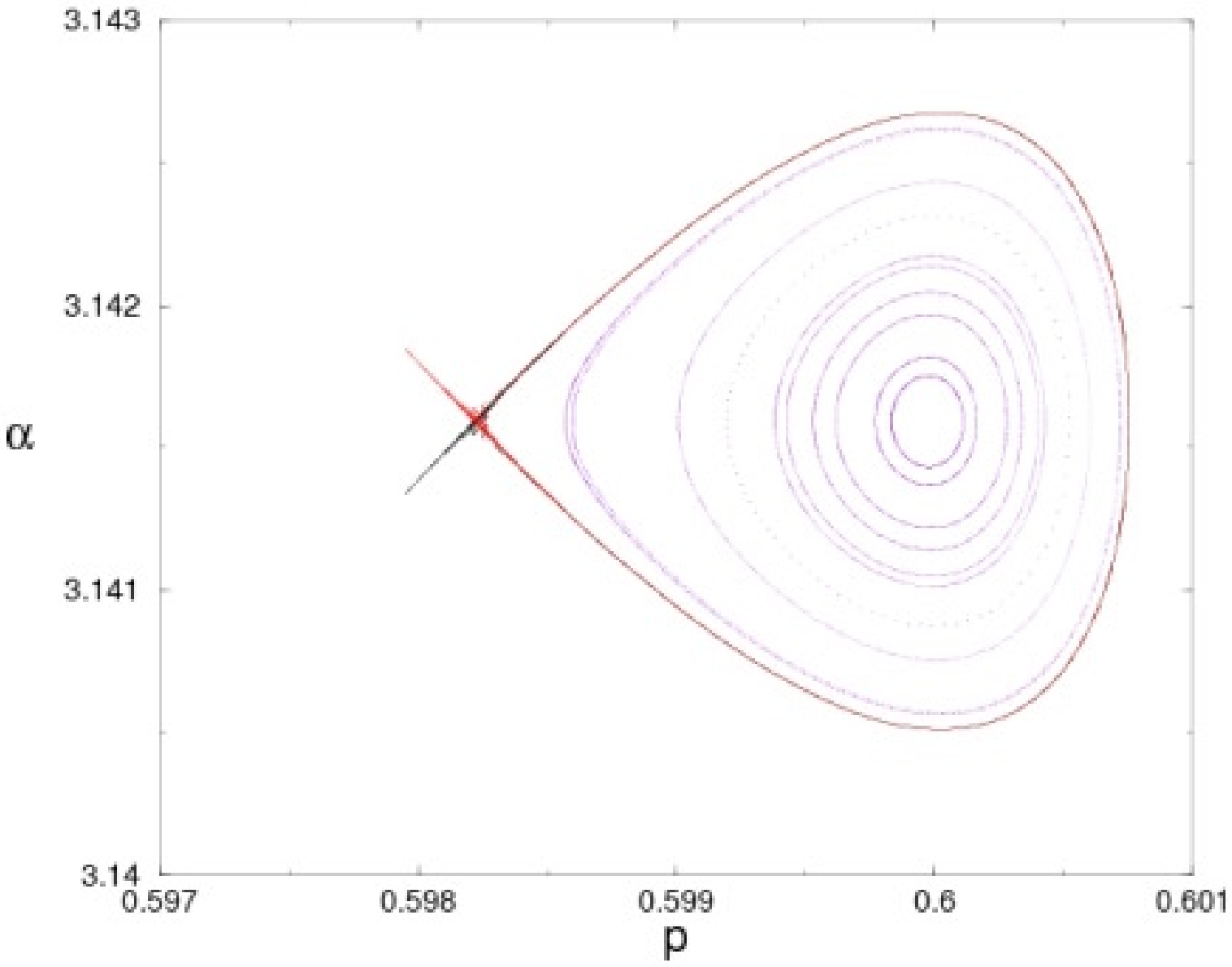} \hspace{0.2cm}
  \includegraphics[angle=270,width=5.5cm]{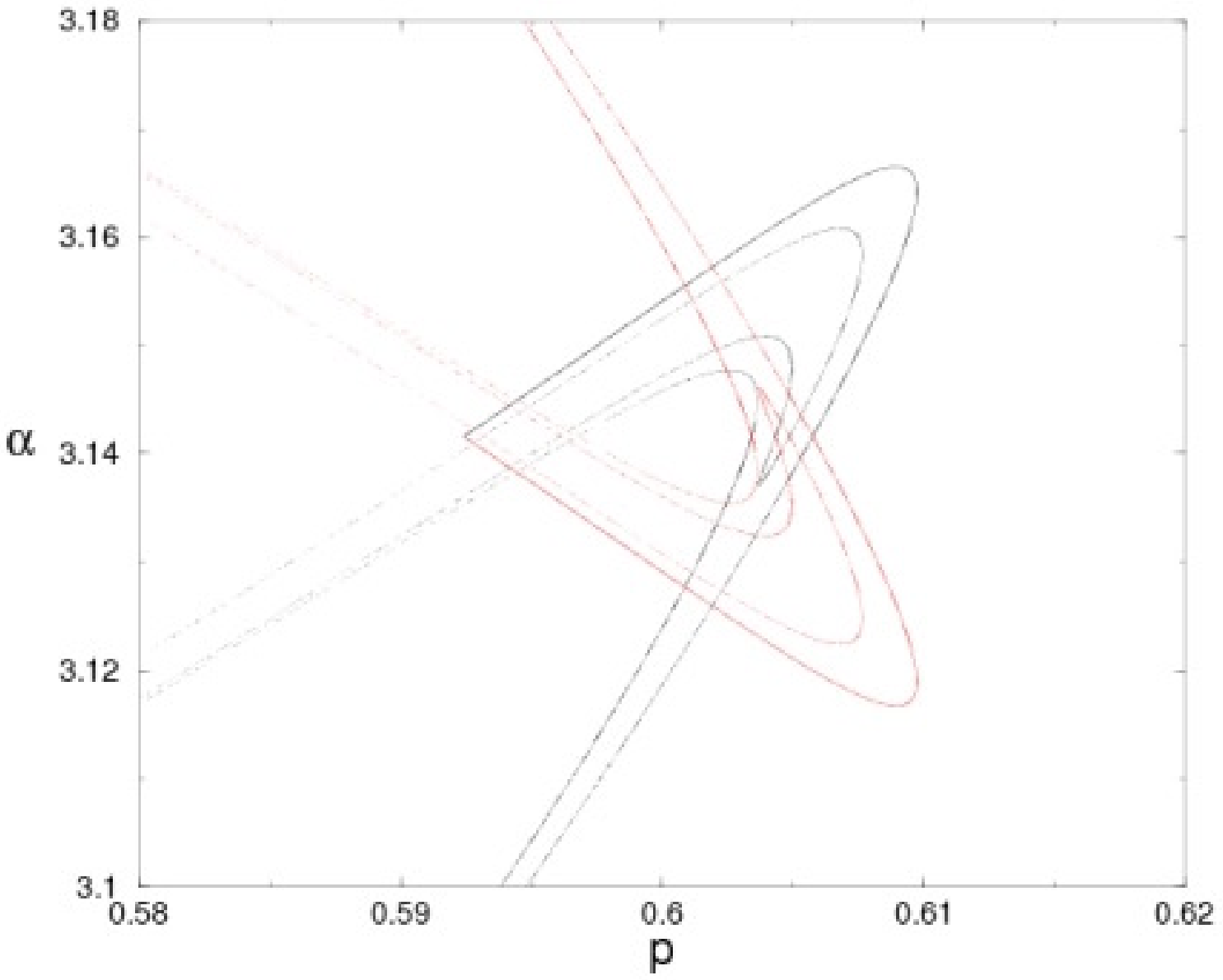}}
\caption{ Phase space structure of the scattering billiard on a
  circular orbit near the maximum of the $n=0$ curve ($R=1$ and
  $d=1/2$). (a)~The horseshoe displays an elliptic and a hyperbolic
  fixed point ($J/(R-d)^2=0.29325$). (b)~Both fixed points are
  unstable and the dynamics is dominated by scattering events, although
  the horseshoe is not yet completely hyperbolic
  ($J/(R-d)^2=0.29218$).}
\label{fig3}
\end{figure*}

\begin{figure*}
  \centerline{ 
    \includegraphics[angle=270,width=9cm]{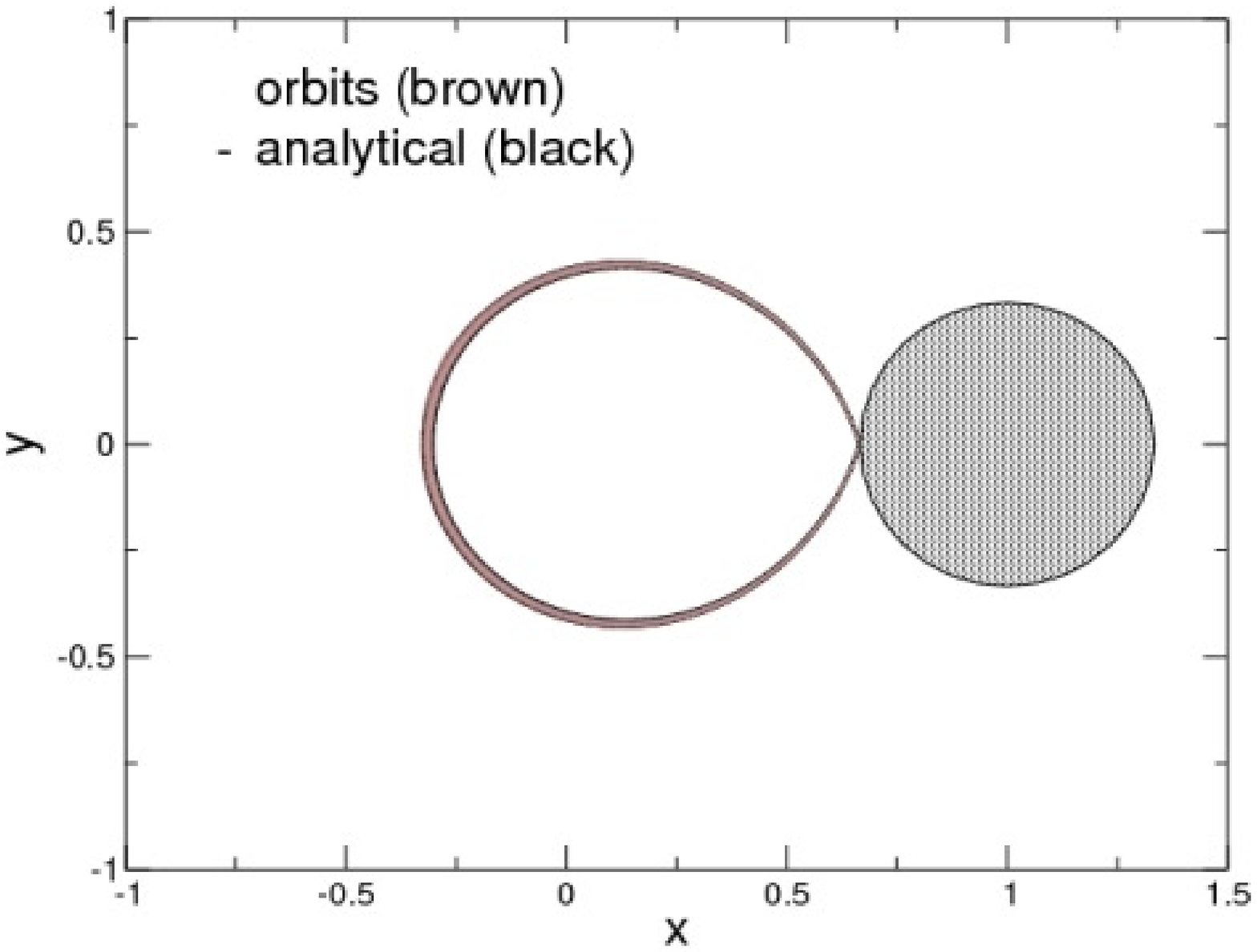} \hspace*{-20pt}
    \includegraphics[angle=270,width=9cm]{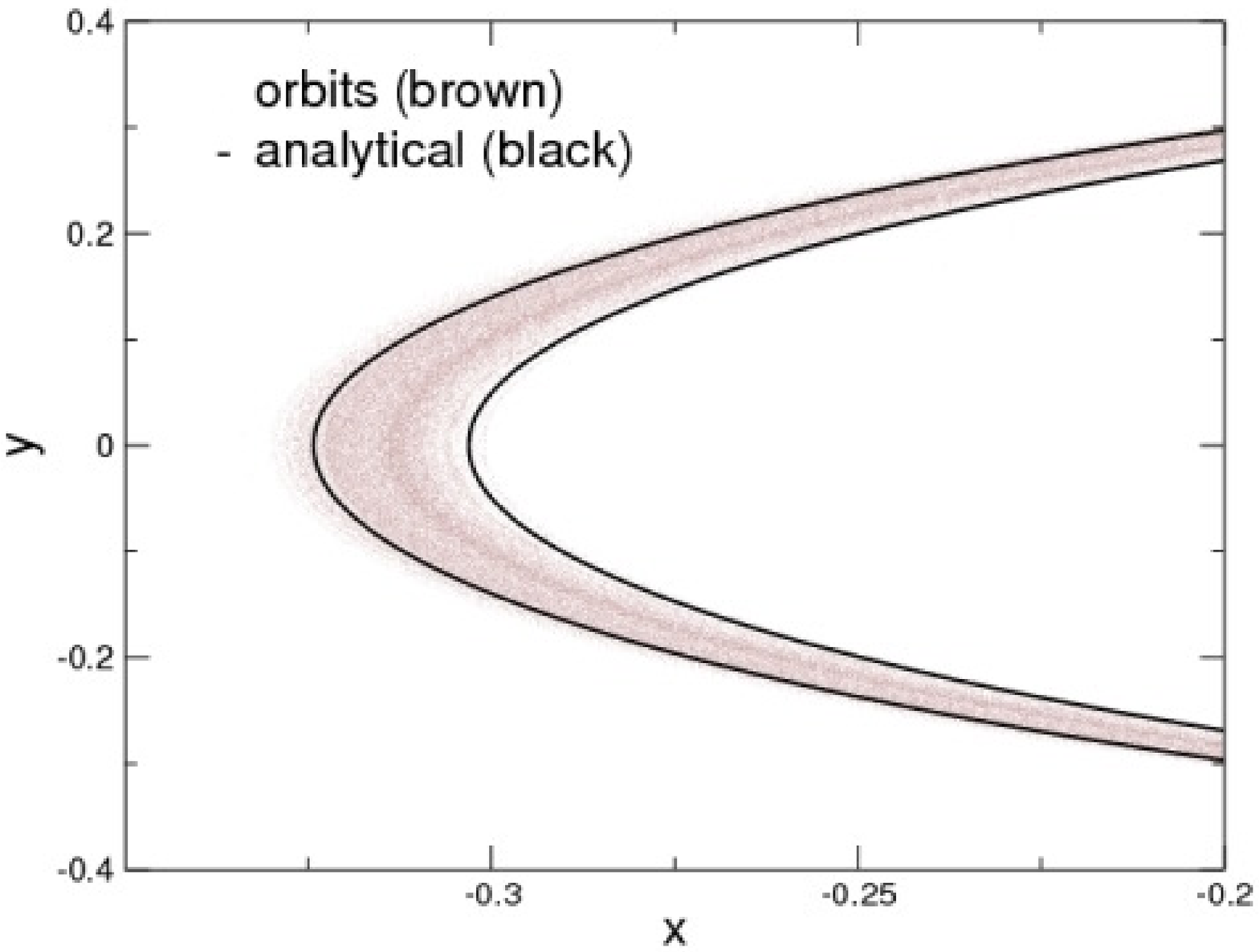}}
  \caption{ (a)~Stable ring of non-interacting particles of the
    scattering billiard system on a circular orbit corresponding to
    the $n=0$ stable periodic orbit. (b)~Zoom of a region of the
    ring. The black lines are the analytical estimates given by ${\rm
      Tr\,} D{\cal P}_J = \pm 2$. }
\label{fig4} 
\end{figure*}

Figure~\ref{fig2}b shows connected intervals in $\theta$, and
therefore in $J$, where the radial collision periodic orbits are
strictly stable, i.e. $| {\rm Tr\,} D{\cal P}_J| < 2$. For $J$ within
these intervals, the phase space structure of the scattering system
displays one elliptic fixed point, surrounded by typical KAM tori and
some chaotic layers (Fig.~\ref{fig3}a).  This region in phase space is
strictly bounded by the stable and unstable invariant manifolds of the
companion hyperbolic fixed point. Particles whose initial conditions
lie within this region display infinite consecutive collisions and are
thus dynamically trapped. Initial conditions that do not belong to any
region of trapped motion eventually escape to infinity. Consider the
case $n=0$ for concreteness. After the appearance of the pair of
periodic orbits (when ${\rm Tr\,} D{\cal P}_J =2$), by further
reducing the value of $J$, there occurs a value where ${\rm Tr\,} D{\cal
  P}_J =-2$. There, the elliptic fixed point becomes inverse
hyperbolic, and a period-doubling bifurcation sets in: the region of
trapped motion vanishes rapidly after further reducing $J$. In
Fig.~\ref{fig3}b we illustrate this case, plotting the
phase space structure corresponding to an incomplete Smale
horseshoe~\citep{RuecJ94}. The dependence of ${\rm Tr\,} D{\cal P}_J
=-2$ upon $d/R$ implies that the actual parameters of the billiard
influence the width of the ring. Physically, this means that the
parameters related with the motion and the mass ratios of the shepherd
moons do indeed influence the width of the ring. Further reducing the
value of $J$ yields a hyperbolic Smale horseshoe.

We turn now to the rings~\citep{Benet00}. Consider an arbitrary large
number of non-interacting particles. Their initial conditions are
completely arbitrary except for the value of their Jacobi integral,
which satisfies the condition $|{\rm Tr\,} D{\cal P}_J| <2$ for the
$n=0$ curve. Then, the particles may collide with the disk at any
value $\phi$ along the circular orbit of the disk. Letting the system
evolve, many particles escape after a few collisions; the remaining
ones are dynamically trapped. Note that the measure of the latter set
is strictly positive, that is, there is a non-vanishing probability of
finding such initial conditions. The motion of the particles belonging
to this set, the ring particles, may be periodic, quasi-periodic or
chaotic. The ring is the pattern formed by these trapped particles in
the $X$--$Y$ plane (inertial frame). Figure~\ref{fig4} shows the
pattern obtained after a few thousand revolutions of the disk,
resulting from a large number of particles whose initial conditions
are close to the maximum of the $n=0$ curve. The plot shows the
$(X,Y)$ position of all particles, in an inertial frame, which have
not escaped at the time when the ``photograph'' was taken.

In Fig.~\ref{fig4} we also plot the curves corresponding to ${\rm
  Tr\,} D{\cal P}_J = \pm 2$, which serve as estimates of the ring
shape. They do not coincide perfectly with the boundaries of the ring,
since they are related to the appearance and destruction of the
central elliptic fixed point and carry no information on the size of
the island around the elliptic points. Yet, the nature of these curves
and the scattering dynamics implies a sharp-edged ring. The narrowness
of the ring is a consequence of the relatively small region, in phase
space, of trapped motion; this also takes into account the (small)
variations induced by changing $J$. Furthermore, the ring can be
characterized as eccentric, in the sense that there are two points,
the periapse and apoapse, forming a line that intersects the origin,
the line of apsides. Finally, we mention that the same ring may be
obtained by an ensemble of particles whose initial conditions, while
different, are very localized in a region in the $X$--$Y$ plane. The
time evolution in this case not only selects the particles of the
ring, but also spreads them over the whole region of trapped motion
(in phase space) on a short time scale.

\begin{figure}
  \centerline{ \includegraphics[width=7.5cm]{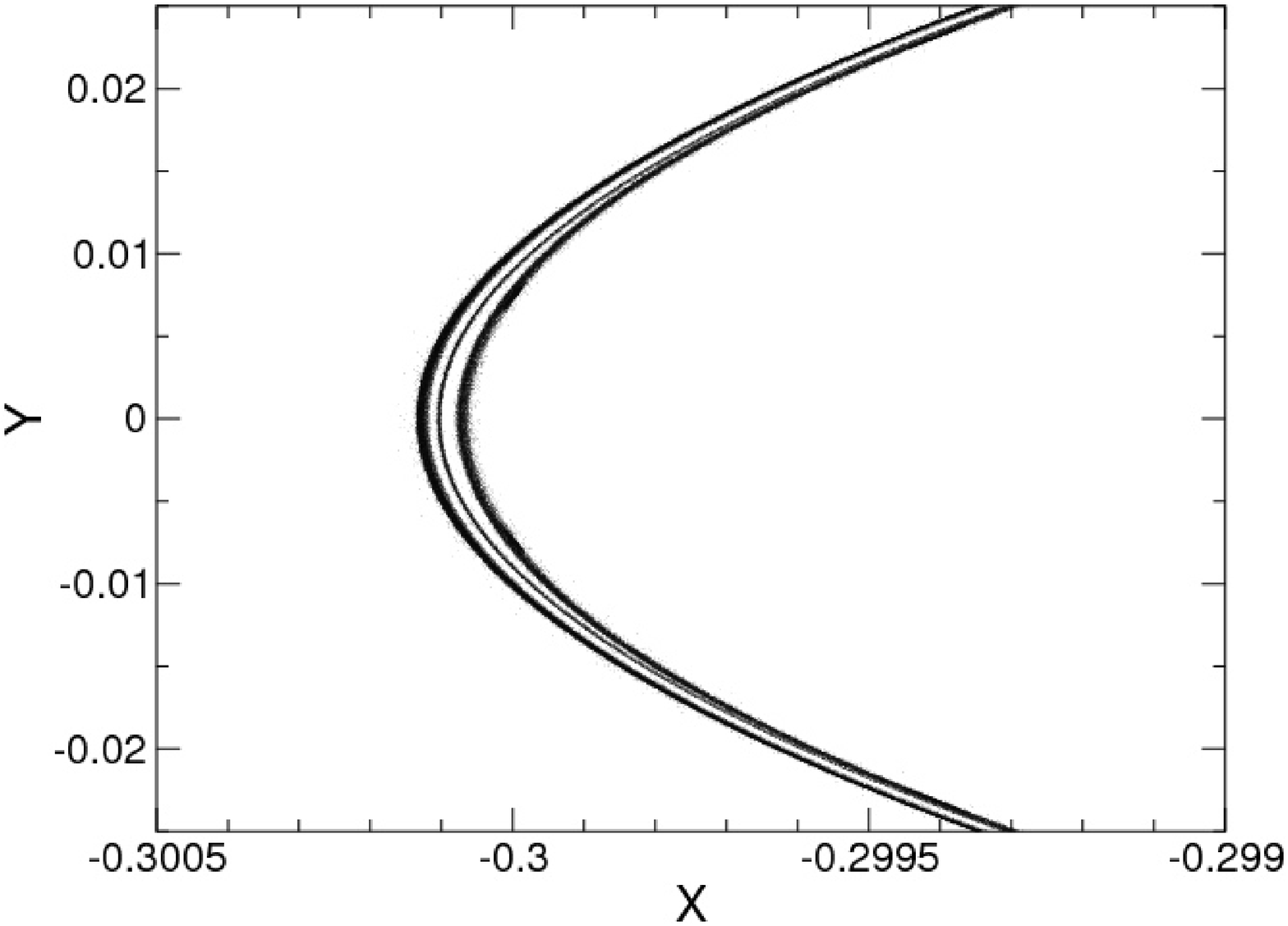}\hspace*{7pt}
    \includegraphics[width=7.5cm]{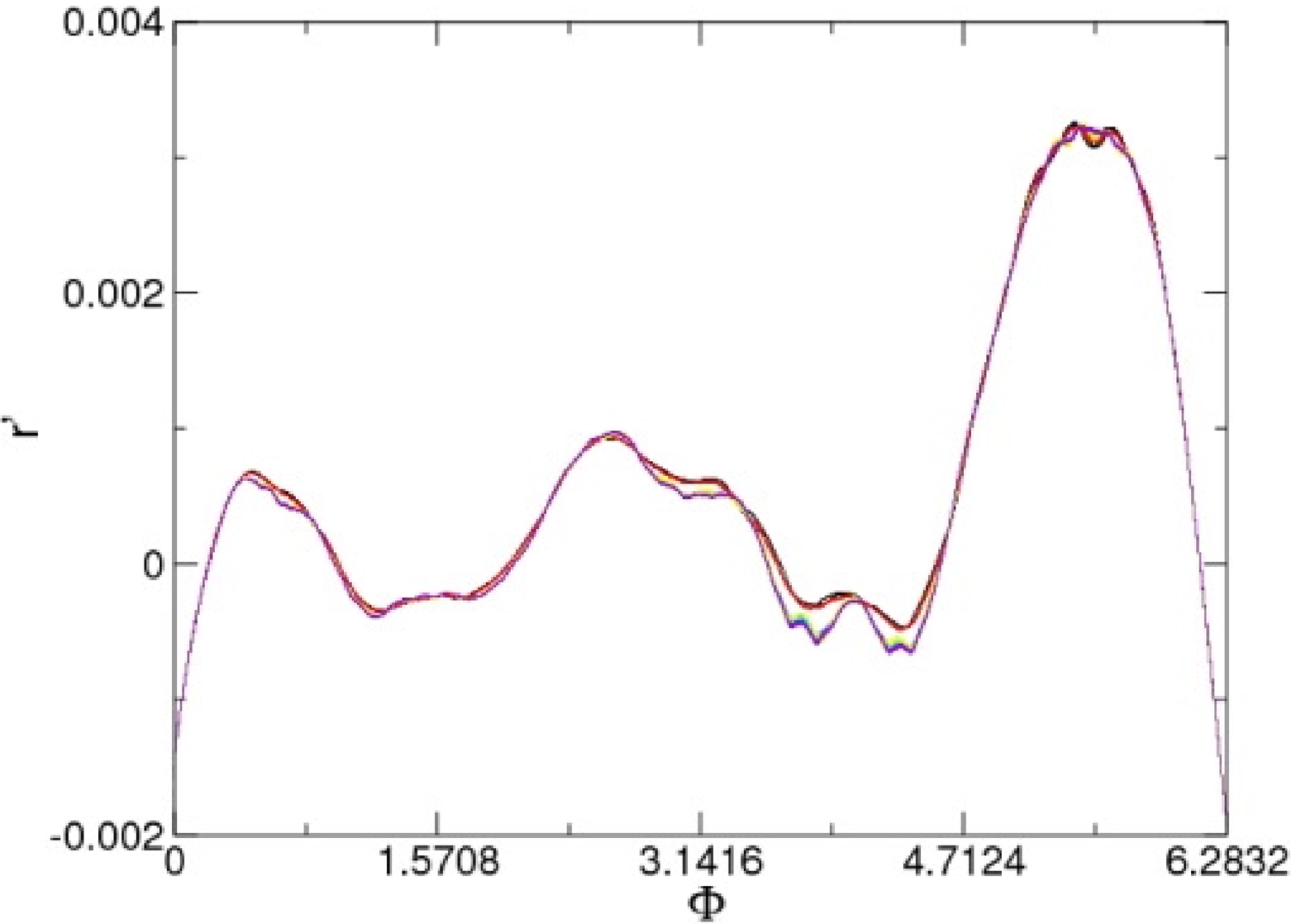} } 
  \vspace{25pt}
  \centerline{ \includegraphics[width=7.5cm]{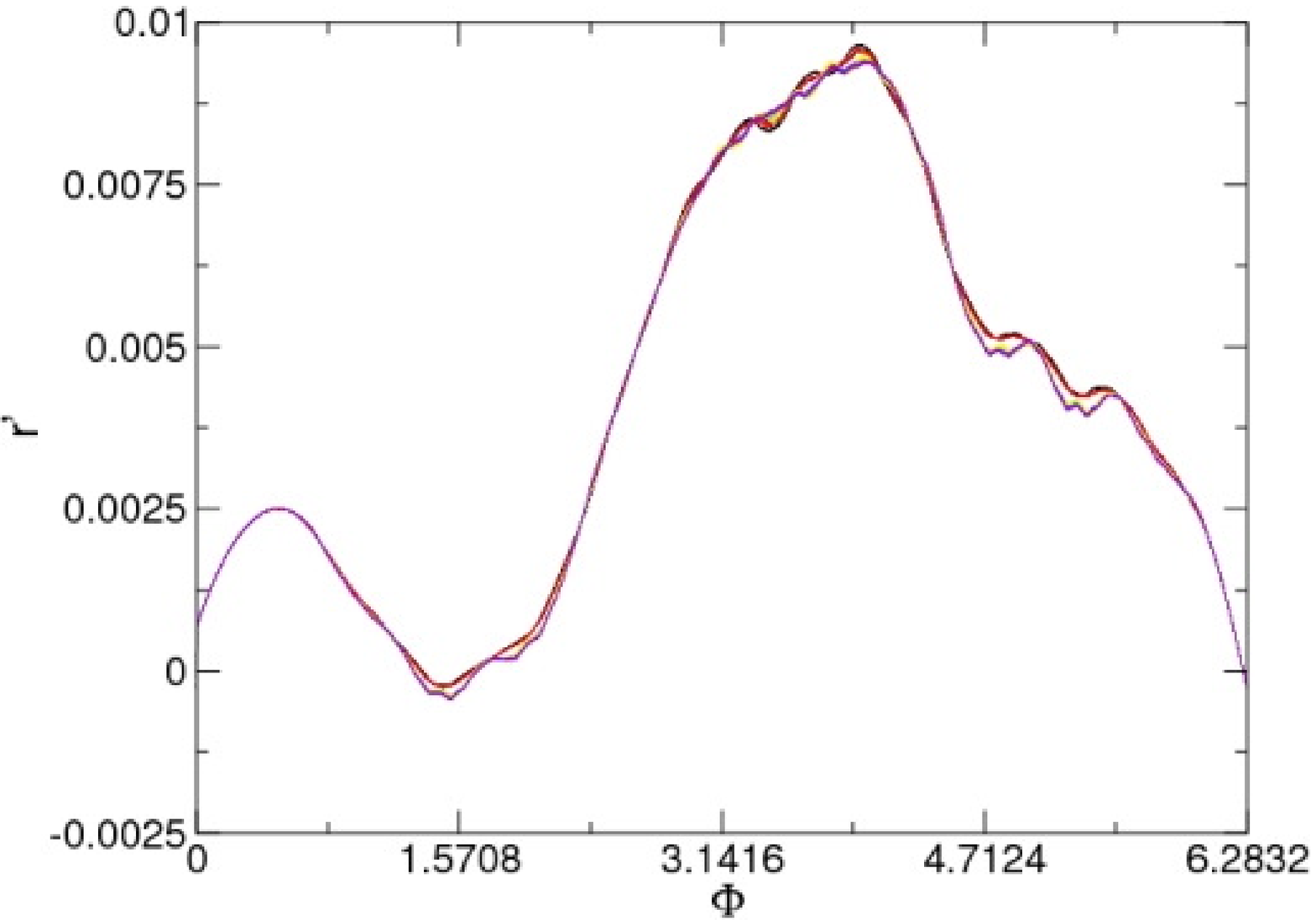}\hspace*{7pt}
    \includegraphics[width=7.4cm]{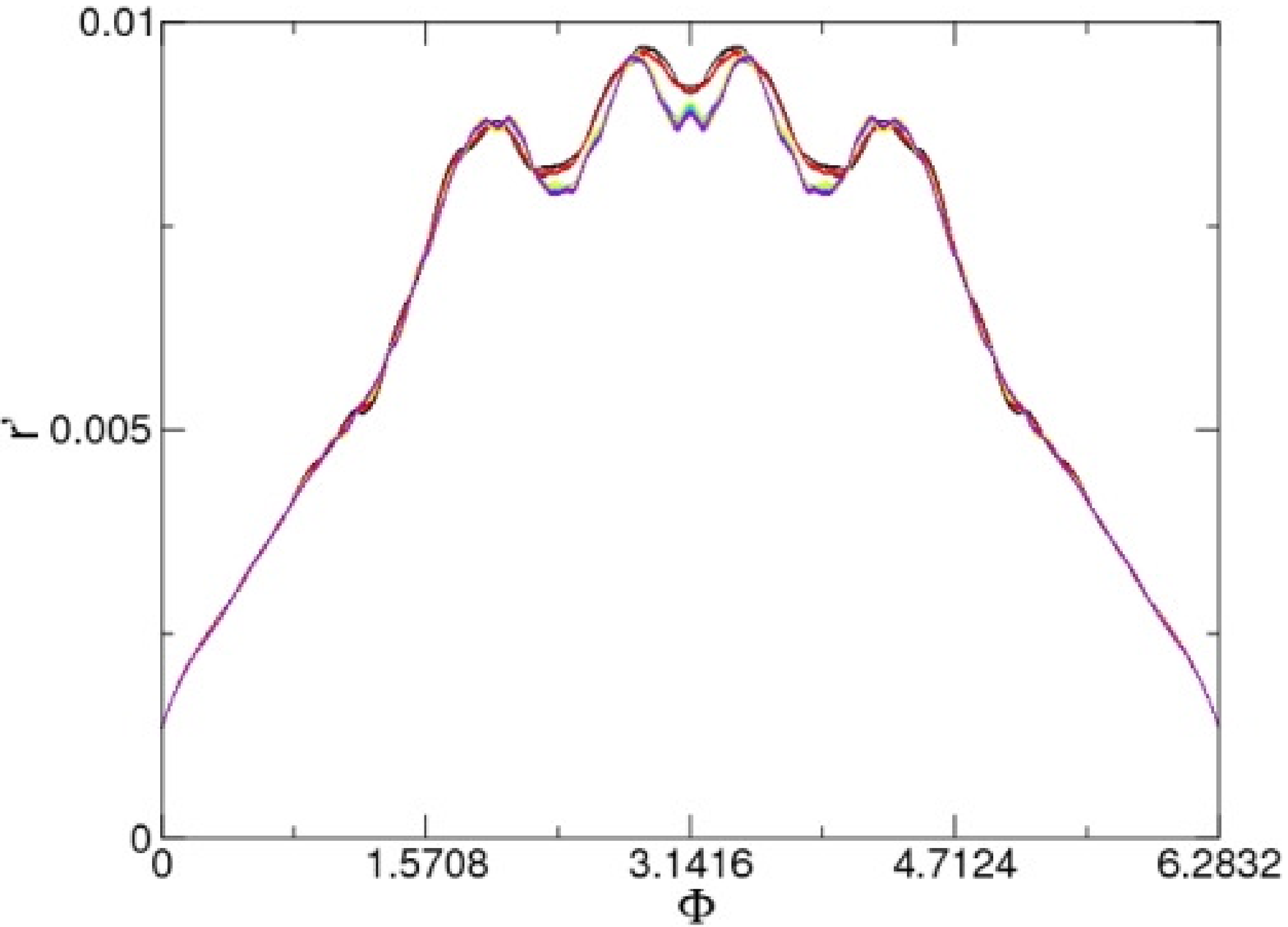} }
  \vspace*{8pt}
  \caption{ (a)~Detail of the ring when the disk moves on an eccentric
    Kepler orbit ($\varepsilon=0.00167$). Whole ring represented in
    terms of $r'$ (see text) for (b)~$t=T_d/10$ (top right), (c)~$t=3
    T_d/10$ (bottom left) and (d)~$t=T_d/2$ (bottom right). Here,
    $T_d=2\pi$ is the period of the disk's orbit.}
\label{fig5}
\end{figure}

\subsection{Rings in the case of elliptic motion}
\label{sec3.2}

In the case of elliptic motion of the disk, Eq.~\RefEq{eq2} is
explicitly time-dependent. The organizing centers in phase space are
now stable tori, which are indeed found at least for small enough
values of the eccentricity.  Close to these stable tori, the structure
of phase space is somewhat similar to that discussed above for
circular motion; details are given in \citet{Benet05}. Hence,
rings occur just as before. For very small values of the eccentricity,
the rings have essentially the same structure as in the circular
case. Increasing $\varepsilon$ slightly rapidly shrinks the width of
the ring. Eventually, we reach a value of the eccentricity for which
the ring displays gaps, i.e., regions of extremely low density of
particles. This is illustrated in Fig.~\ref{fig5}a. In this case
($\varepsilon=0.00167$), the gaps divide the ring into three
components, which we call strands. They are a direct consequence of
the stable tori that yield the ring, which arise from the
quasi-periodic character of $V_1$ in Eq.~\RefEq{eq1}. We observe that
particles belonging to a given strand stay in the same strand, i.e.,
they do not move along other components of the ring.

In Fig.~\ref{fig5}b we show a different representation of the whole
ring for a specific time (given as a fraction of the disk's period
$T_d$). We plot the azimuthal position $\Phi$ of the ring particles,
measured anti-clockwise from the collision point with the disk as
origin, and $r'=r-r_{\rm fit}$, where $r$ is their radial position and
$r_{\rm fit}$ an appropriate fit to the mean value. This figure shows
clearly the intricate entanglement of the strands, reminiscent of the
beautiful braids of Saturn's F ring. We emphasize that the different
strands and braids we observe are consequence of the fact that a
stable quasi-periodic torus is now the organizing center of the
trapped region in phase space. Figures~\ref{fig5}c--d display the ring
at other fractions of the period of the disk. Together these plots are
a representation of the orbital motion of the ring, i.e., the
dynamical behavior of the braids.

We emphasize that the strands and braids we have obtained for this toy
model, which display a dynamical evolution, are not short-lived
structural properties. Thus, stable tori allow for the occurrence of
structured narrow rings. Moreover, and as we shall show in the next
sections, the ideas described here can be extended to more realistic
situations. This is so because of the robustness of relying on
considerations in phase space. The above description of the appearance
of strands and braids should therefore hold in more realistic
situations.

\section{Example 2: Rotating billiard with a central gravitational
  interaction}
\label{ex2}

In the previous section we illustrated, using a billiard system, our
approach to the occurrence of narrow rings and how some observed
structural properties arise. This example, while showing the
robustness of the ideas of the approach, constitutes the simplest
scattering system that is able to produce narrow rings, and as such
has a pedagogical value. However, it lacks of the gravitational
potential with the central planet. In this section we include a
central attractive $1/r$ interaction. As we shall show below, the
present case establishes the connection to the gravitational
restricted three-body problem, which we consider in Sec.~\ref{rtbp}.

Consider the billiard system on a circular orbit with the addition of
a central attractive gravitational interaction, i.e., $V_0=-1/r$
between the (ring) particle and the central planet. In distinction to
the pure billiard, where the free motion is rectilinear and uniform
(the potential is $V_0=0$), the particles now move along solutions of
the Kepler two-body problem. Besides such (collisionless) solutions,
there are also trajectories which collide with the disk. We focus on
the latter, as they model rings close to shepherds; the former would
trivially form a two-body keplerian ring, which is circular and
wide. The collision solutions are constructed by segments of keplerian
trajectories, glued together at the collision points on the disk,
where the precise outcome of the collisions is taken into account.

We proceed as before and compute the consecutive-collision periodic
orbits in the rotating frame. To this end, we observe that the radial
collisions conserve, as before, energy and angular momentum, and occur
only at the innermost ($r_{-}=R-d$) or outermost ($r_{+}=R+d$) points
of the disk. As mentioned above, in the present case we must also
account for the keplerian {\it free flight} between the
collisions. The present problem is thus equivalent to finding periodic
orbits in a central gravitational field which collide consecutively
with a moving point singularity on a circular orbit of radius $r_{-}$
and $r_{+}$, respectively. This singularity influences the motion of
the third particle only through collisions. This is so as long as the
trajectory avoids other collisions with the disk; if not the periodic
orbit is simply destroyed. We note that this is essentially the same
as the circular RTBP with $\mu=0$, as defined by \citet{Henon68}.

These statements can be formulated quantitatively as follows. The
solutions of the keplerian free flight are precisely conic
sections. We consider for concreteness the elliptic motion. Using
symmetry arguments, it is easy to show that consecutive periodic
orbits can be obtained from initial conditions (of the moving
singularity and the third particle, see~\citealt{Henon68}) on the
$X$-axis of the inertial frame. This axis is parallel to the major
axis of the ellipse. The elliptic motion of the third particle, in an
inertial frame, is parameterized by
\begin{eqnarray}
X_p &=& \sigma_0 a_e(\sigma_2 \cos E - \varepsilon_e),\\
Y_p &=& \sigma_0 \sigma_1 \sigma_2 a_e (1-\varepsilon_e^2)\sin E,\\
t &=& a_e^{3/2} (E-\sigma_2\varepsilon_e\sin E).
\label{eqs-ellipse}
\end{eqnarray}
Here, $a_e$ and $\varepsilon_e$ are respectively the major semi-axis
and eccentricity of the elliptic orbit, and $E$ is the eccentric
anomaly. The parameter $\sigma_0$ defines if the periapse lies on the
positive ($1$) or negative ($-1$) $X$-axes, $\sigma_1$ determines if
the trajectory is direct ($1$) or retrograde ($-1$) and $\sigma_2$
whether the initial condition is at periapse ($1$) or apoapse
($-1$). Then, in order to obtain the periodic orbit one needs to match
$(X_p,Y_p)$ with $(r_\pm\cos t,r_\pm\sin t)$ and solve for $a_e$ and
$\varepsilon_e$. With $\sigma=\sigma_0\sigma_2$, $\sigma_\eta={\rm
  sgn}(\sin\eta)$ and $\rho = \sigma_\eta (1 -
\sigma\cos\tau\cos\eta)^{1/2}$, the timing condition reads
\begin{equation}
\label{eq_mu0}
F_0(\tau,\eta;\sigma,\sigma_\eta) = \rho
[\eta\rho^2-\sin\eta(\cos\eta-\sigma\cos\tau)]-
\frac{\tau\sin^3\eta}{r_\pm^{3/2}} = 0.
\end{equation}
\citet{Henon68} computed first the solutions of~\RefEq{eq_mu0}
for $r_+=r_-$ ($d=0$).

\begin{figure}
  \vspace*{10pt}
  \centerline{
    \includegraphics[width=5.5cm]{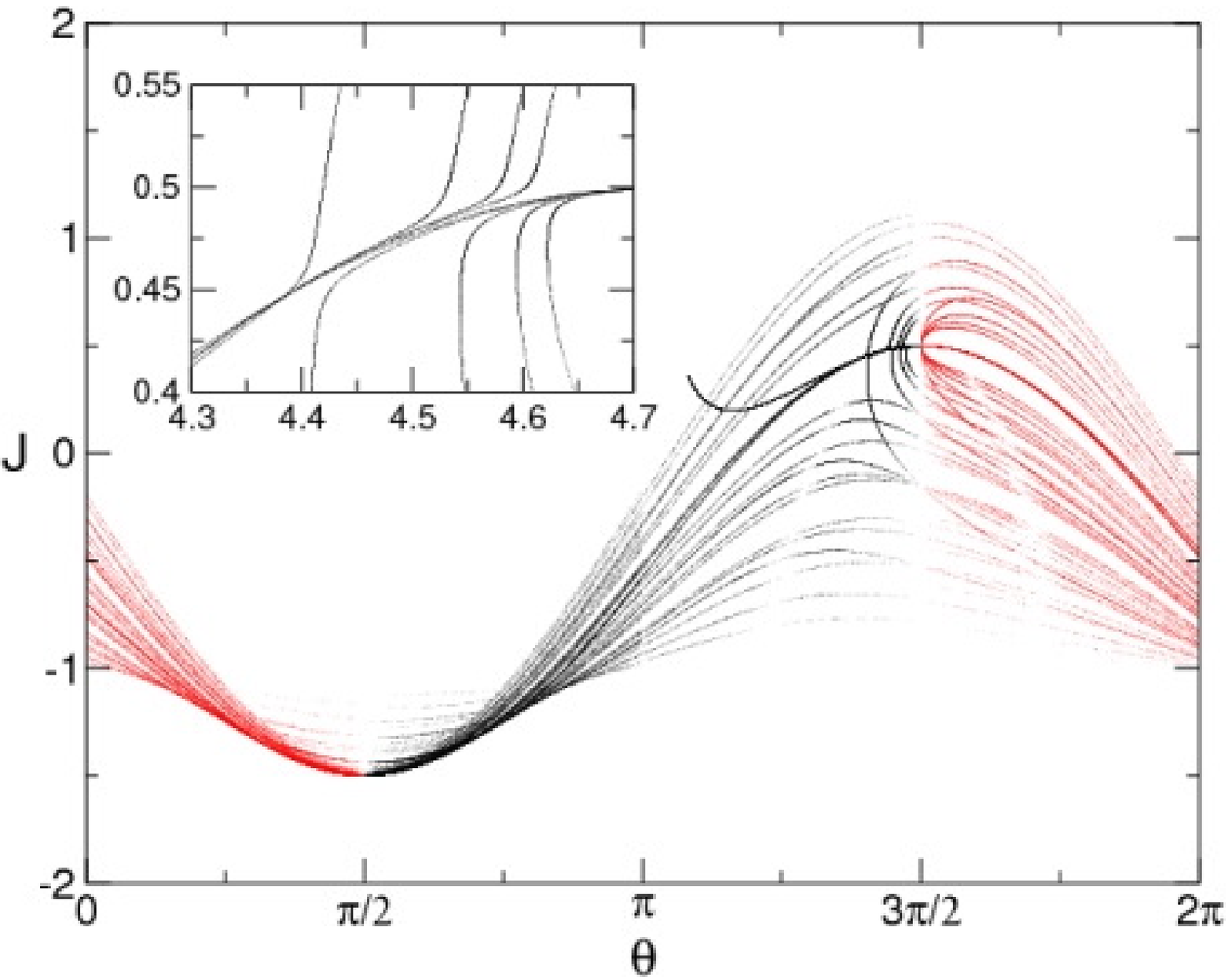}
    \hspace*{.5cm}
    \includegraphics[width=5.5cm]{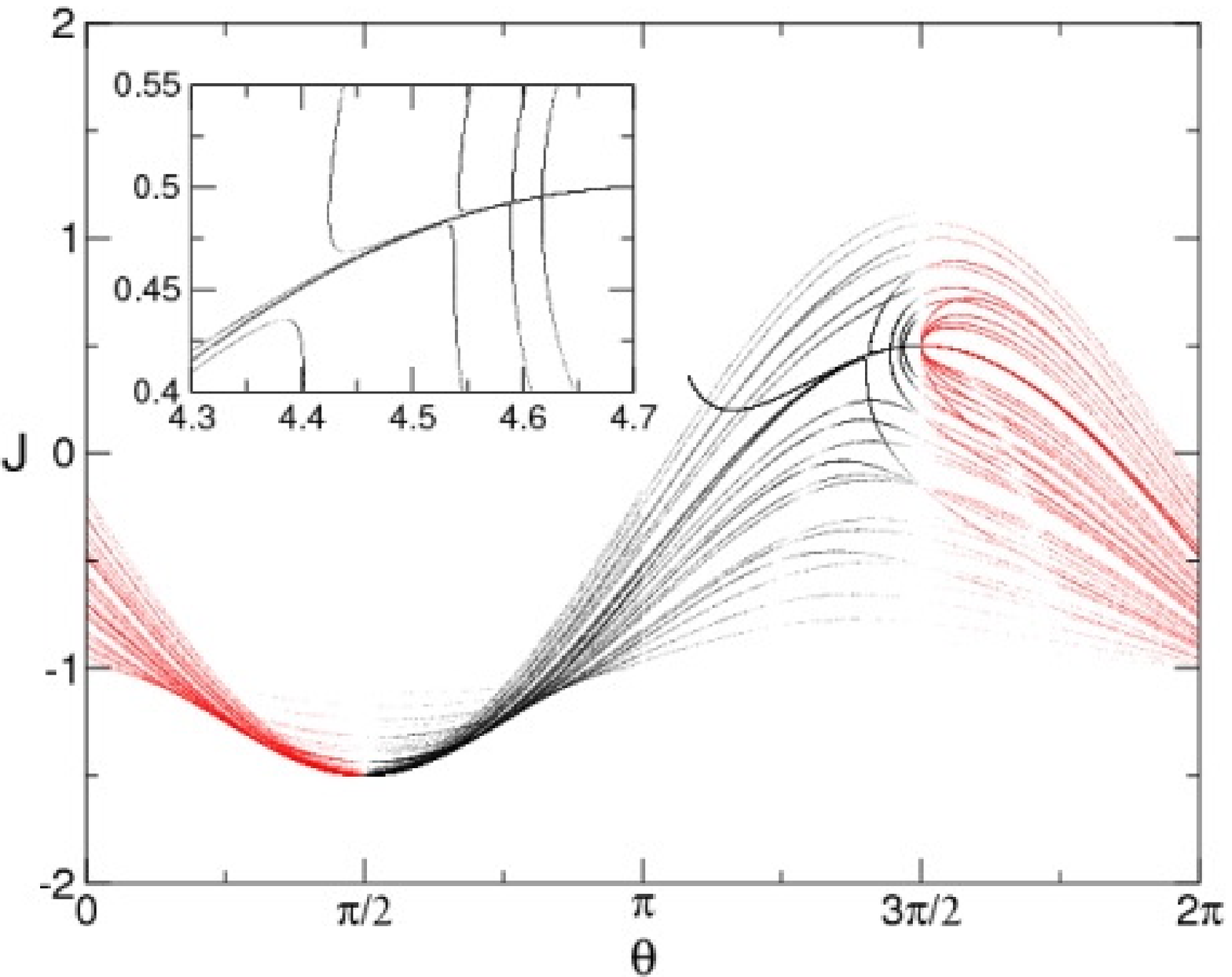}
    \vspace*{10pt} }
  \caption{ Periodic orbits of the disk on a circular orbit with a
    central gravitational interaction for (a)~$d=10^{-3}$ and,
    (b)~$d=0$. The insets show differences between the two cases.}
  \label{fig6}
\end{figure}

In Fig.~\ref{fig6}a we show the consecutive collision periodic orbits
for a disk of radius $d=10^{-3}$; in Fig.~\ref{fig6}b the case $d=0$
is plotted for comparison. At first sight, no difference can be
observed. The case $d=0$ corresponds precisely to the circular RTBP
with mass parameter $\mu=0$~\citep{Henon68}. This naive remark has
important consequences. We emphasize that we are interested not in the
billiard problem with a central gravitational potential, which is only
another example, but on the full gravitational problem, with its
simpler version, the RTBP. \citet{HitzlHenon77} proved that the
limit, in the case of vanishing $\mu$, of the second species periodic
orbits of the circular RTBP with small but non-zero $\mu$ are the
{\it critical points} of the consecutive collision periodic orbits for
$\mu=0$ (Fig.~\ref{fig6}b). This statement proves the validity of our
approach for the gravitational circular RTBP, and establishes the
connection of our naive examples to the gravitational case. Using the
arguments of Section~\ref{theory}, this validity can be extended to
problems with other interactions as well as more degrees of freedom.

Coming back to the billiard with the central gravitational
interaction, we are interested in the regions of phase space where
scattering may take place. In this example, scattering is
possible for positive keplerian energy. In the $\theta-J$ plane, 
scattering motion occurs if the Jacobi integral satisfies, in
some $\theta$ interval, the inequality $J \ge \max(J_{-},J_{+})$, with
$J_{+} = -(2 r_0)^{1/2}\sin\theta$ and $J_{-} = -1 / r_0$
($r_0=r_\pm$, depending on the value of $\theta$). We notice that for
$J>0$ scattering is possible.

To construct the rings, we proceed as before, requiring that the
periodic orbits where scattering is achievable are, in addition,
stable. Knowing the periodic orbits it is easy to compute their
stability. The same procedure then yields the rings, though some
subtleties arise. These are related to the fact that the particle's
energy must be positive for the particle to escape. The only way of
changing its energy is through collisions with the disk (which is why
non-colliding orbits form trivially a wide circular ring). However,
the change in the energy may not be so large as to lead to escape
after a single collision, so that repeated collisions are typically
required for escape. Since between collisions the particle moves along
a Kepler ellipse, and we are interested in a tiny disk, the time
between consecutive collisions can be rather long. In a Poincar\'e
surface of section, defined when the particle crosses the $x$ synodic
axis (e.g, $x$ vs. $p_x$), such long motions along a Kepler ellipse
will be represented as something resembling a torus. The reason to
consider such a Poincar\'e surface of section is inspired by the
RTBP. However, these are {\it fake tori}, not true tori, since collisions
will abruptly change the particle into a different one (if the energy
is still negative) or lead to escape. Therefore, long integrations
(about $10^5$ periods of the disk or even longer) are required in
order to distinguish whether the trajectory is truly trapped, and not
to ``misguide the eye'' with the fake tori described above.

\begin{figure}
  \centerline{
    \includegraphics[width=5cm,angle=270]{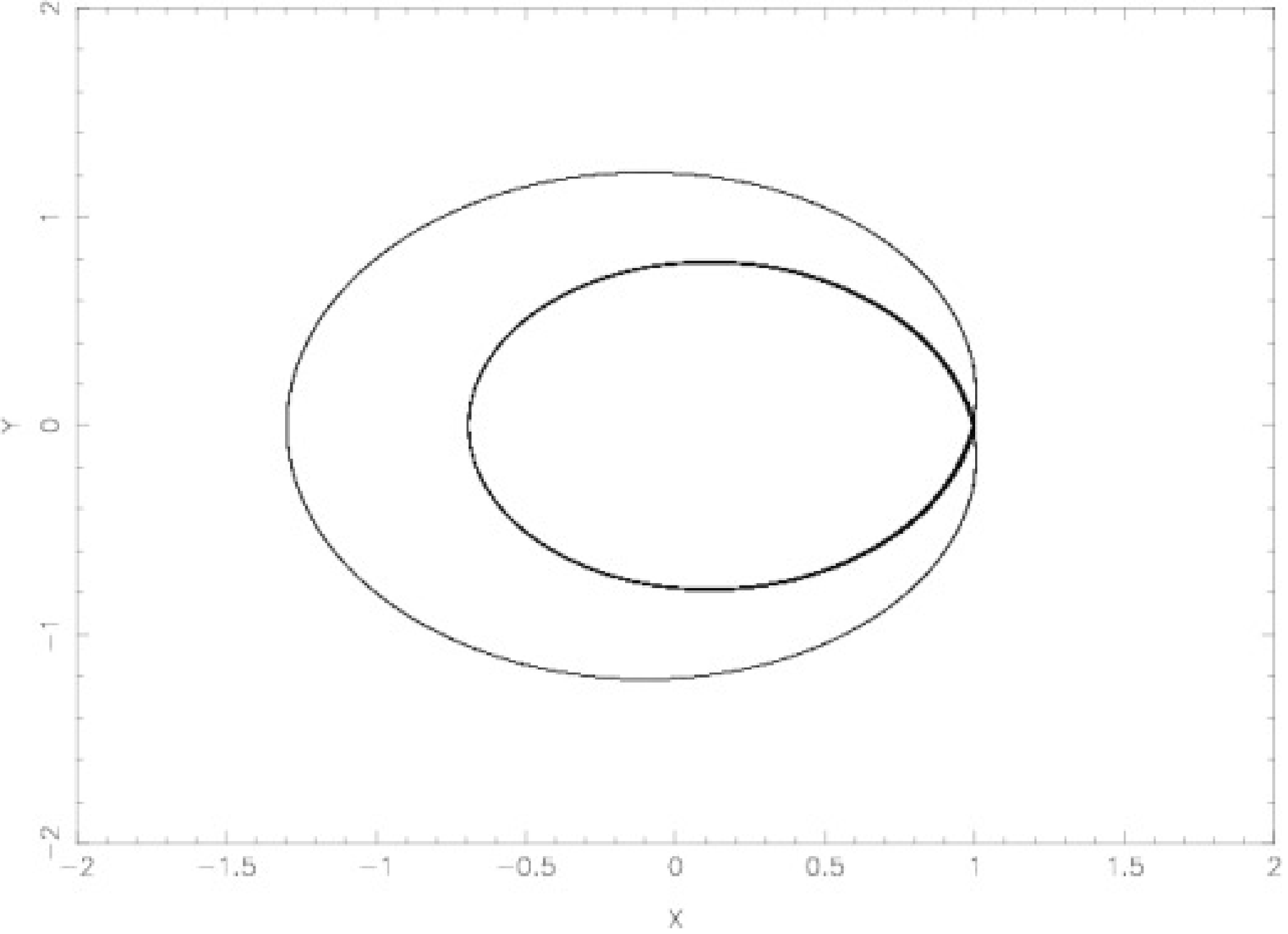}\hspace*{7pt}
    \includegraphics[width=5cm,angle=270]{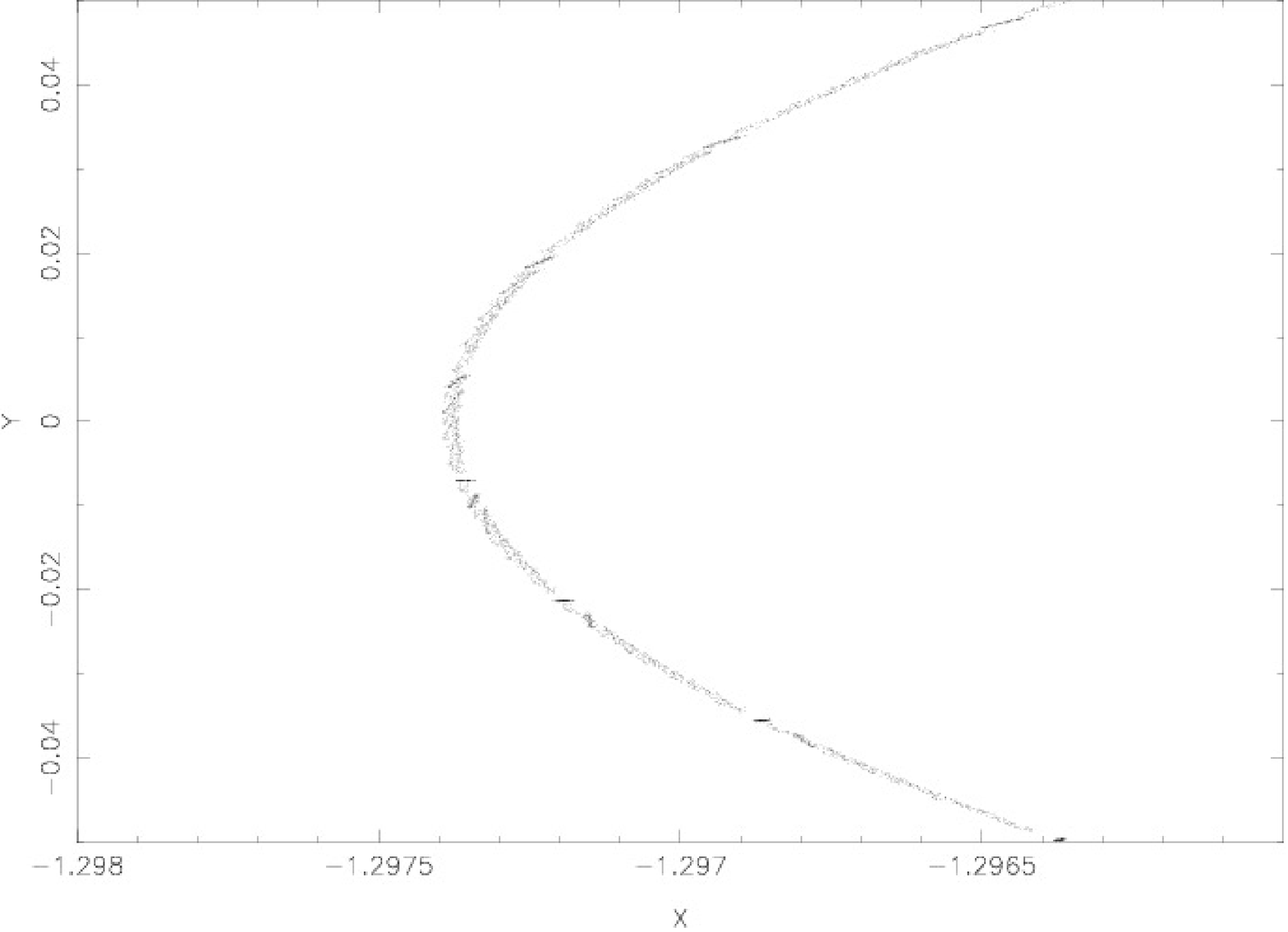}
    \vspace*{10pt} }
  \caption{ (a)~Ring generated by the $C_{12}(2)$ periodic orbit for
    the circular rotating billiard system with a $1/r$ central
    interaction. The multiple components of the ring are a consequence
    of the structure of the organizing periodic orbit, which is closed
    after a few full turns of the disk. (b)~Detail of a region of the
    ring.}
\label{fig7}
\end{figure}

In Fig.~\ref{fig7}a we present the ring generated by the stable
collision periodic orbit, denoted as $C_{12}(2)$ by
\citet{Henon68}, for a disk of radius $d=0.0006$. The disk radius
was chosen as large as possible but such that the characteristic curve
in Fig.~\ref{fig6}a displays a local maximum and therefore the
saddle--center scenario that yields one stable and one unstable fixed
points takes place. As observed in Fig.~\ref{fig7}a, the ring displays
multiple components through the inner loop, reminiscent of a strand
structure, as shown in Fig.~\ref{fig5} for the rotating billiard on an
elliptic orbit. The observed structure is inherited directly from the
shape of the organizing periodic orbit in the synodic (rotating)
frame; there, the organizing periodic orbit is closed after a few full
turns of the disk. Figure~\ref{fig7}b shows an enlargement of a region
of the ring. As expected, the rings are narrow and eccentric, and
display sharp edges.

\section{Example 3: The circular restricted three-body problem}
\label{rtbp}

We now present results of our approach to the occurrence of narrow
planetary rings in the circular restricted three-body problem. We
concentrate on the case where the mass of the secondary (shepherd) is
$10^{-6}$, in units of the mass of the central planet,
i.e. $\mu\approx 10^{-6}$. While this value is much larger than the
actual masses of the known shepherd satellites, it yields qualitatively
the same results with an enormous saving in computing time.

The advantage of considering the rotating billiard disk with a central
$1/r$ potential is that we know the precise location of the
consecutive collision periodic orbits. Moreover, its finite size and
the curvature of the disk enhance the instabilities which finally lead
to escape. In the case of the RTBP, even for very small mass
parameters, this knowledge has to be obtained numerically, despite the
fact that H\'enon's work for the case $\mu=0$ gives a starting
point. The numerics are nevertheless non-trivial, since the orbits of
interest are quite close to collisions with the secondary, which not
only complicates the numerics but requires a rather high level of
accuracy. As we shall see below, we also require rather long
integrations.

We shall focus on {\it non--collision} orbits generated by {\it the
  continuation} of the $C_{12}(2)$ consecutive collision periodic
orbit of H\'enon for $\mu=0$. This illustrates that indeed the
rotating billiard with a $1/r$ central attractive interaction provides
the link between the naive rotating billiard system and the
RTBP. Moreover, this also illustrates the result of
\citet{HitzlHenon77} that the generating periodic orbits for
$\mu=0$ are the second species periodic orbits for finite but small
values of $\mu$. We emphasize that, the organizing centers and the
trapped orbits we consider, and therefore the ring, {\it do not}
display collisions with the secondary. We first use Fig.~\ref{fig6}b
to locate approximately the Jacobi integral interval where the
continuation of the $C_{12}(2)$ periodic orbit is stable. From here,
numerically, we obtain the interval where the periodic orbit has a
stable branch. We find that such a stable orbit exists for $0.400 \le
J \le 0.453$. Note that these values of the Jacobi integral are much
larger than the associated with the zero-velocity
curves~\citep{BST98}. Therefore, scattering dominates the dynamics
unless the proposed mechanism (existence of stable periodic orbits)
creates regions in phase space of dynamically trapped motion. That is,
escaping trajectories largely fill phase space for these values of
$J$, except when trapping occurs near a stable periodic orbit. Below,
we shall focus in the dynamics close to, and including, one specific
region of bounded motion. For the RTBP with very small values of $\mu$
the dynamics consists of almost-elliptic keplerian motion, that takes
place during long times, followed by an eventual close approach that
finally leads to escape~\citep{BST98}. Therefore, the numerical
simulations must distinguish the fake tori described in the previous
section, which again requires long and accurate integrations. In our
simulations we have checked for about $10^6$ periods of the secondary
that no abrupt tori exchange takes place.

\begin{figure}
  \centerline{
    \includegraphics[width=5cm,angle=270]{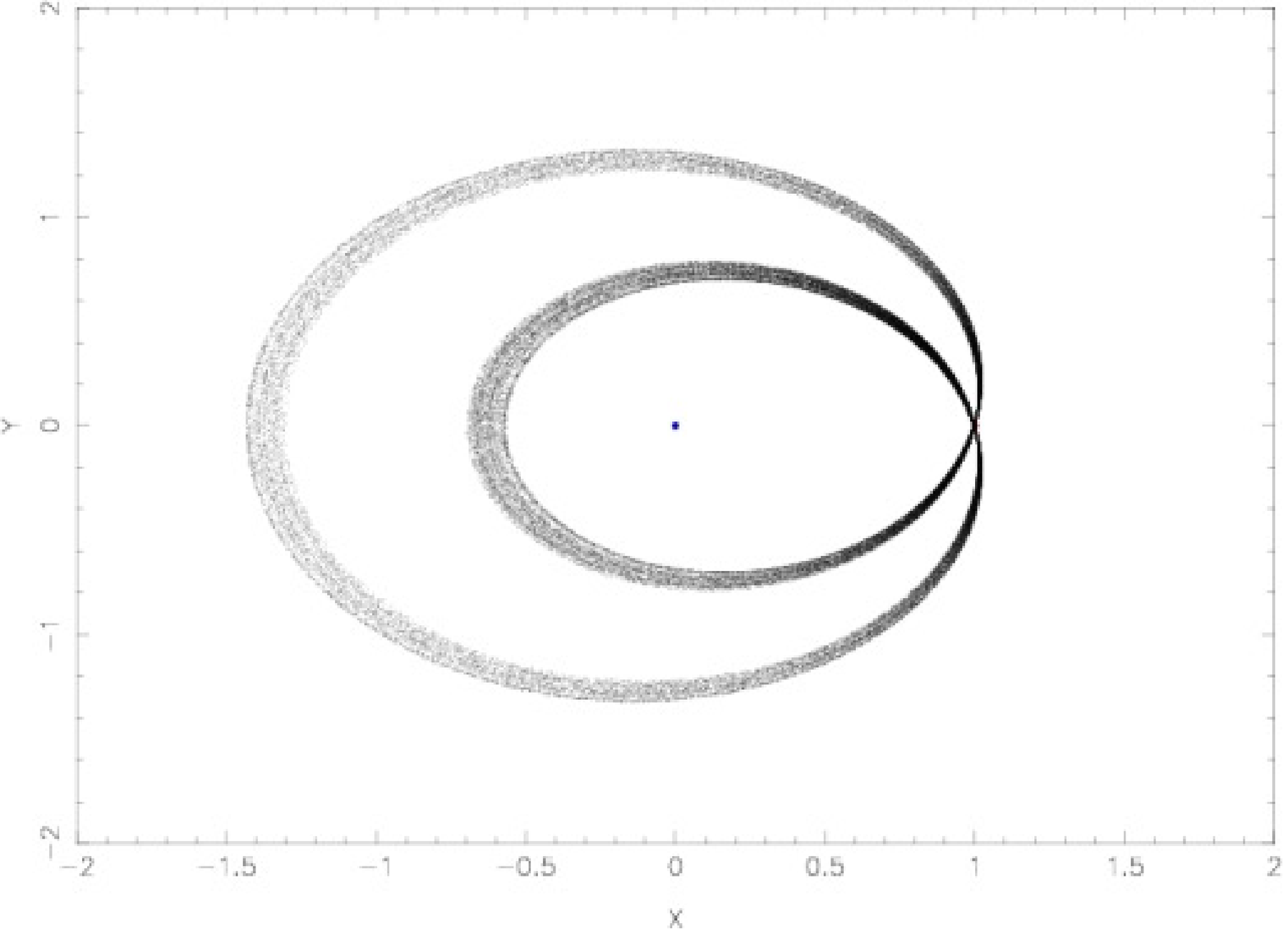}\hspace*{7pt}
    \includegraphics[width=5cm,angle=270]{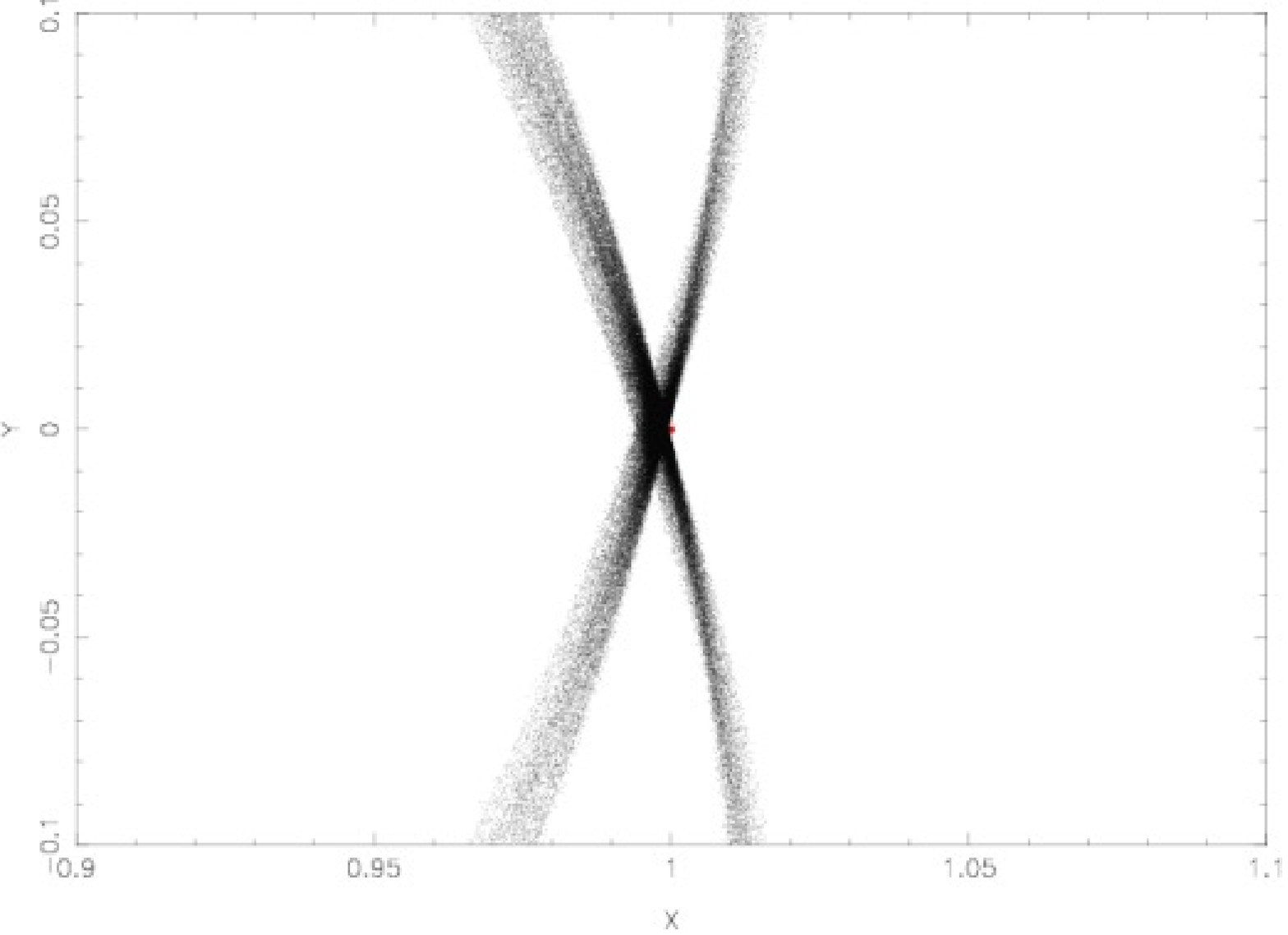}
    \vspace*{10pt} }
  \caption{ Ring generated by the $C_{12}(2)$ periodic orbit for the
    circular RTBP with $\mu\approx 10^{-6}$. (a)~Full view;
    (b)~details of a specific region close to the shepherd. Note the
    dependence of the width with respect to the proximity with the
    shepherd. The shepherd and the central planet are denoted by
    dots.}
\label{fig8}
\end{figure}

Once the stability region (bounded motion) around the periodic orbit
is located and also the approximate interval where it is stable, the
construction of the ring is straightforward, following the ideas we
have described above. Figures~\ref{fig8} show the whole ring and an
enlargement of a region. All qualitative features described in the
previous sections are observed once again: sharp edges and a narrow
eccentric ring with many components. Again, these many components are
inherited from the shape of the organizing periodic orbit in the
synodic frame, which closes after a few full turns of the
disk. Another important observation is that there is a variable width
of the ring, with respect to the position closest to the
shepherd. This property is observed in real narrow planetary
rings~\citep{Esposito02}. The explanation within our framework is
simple: the region of trapped motion is very small close to the
shepherd, for any value of $J$; away from it, the two-body Kepler
interaction allows for longer excursions due to the different values
of $J$ involved. The projection onto the $X$--$Y$ plane is wider as we
move away from the neighborhood of the shepherd.

We note that, for the interval of the Jacobi constant considered,
other regions of dynamically trapped motion exist in phase space. In
particular, there is a comparatively large stable region associated
with a period-one fixed point, and another one related to a period-two
fixed point, both belonging to phase space regions where scattering
dominates the dynamics. From our construction, these regions would
also yield rings. Yet, these stable regions are extremely robust, in
the sense that they are present for an extremely wide interval of
$J$. Focusing on the period-one stable region, it essentially sweeps
out every value of $x$ in the synodic frame. Therefore, except for a
tiny gap which is opened by the shepherd, it yields a wide circular
ring; this ring is of no interest in the context of the present
work. It is the result of the dominance of the central $1/r$ keplerian
interaction. This result implies that the RTBP is not a realistic
model, at least to describe the Uranus rings or the narrow rings of
Saturn. We believe that a realistic model should not only consider the
planet, a shepherd and the ring particle as we have done, but at least
another body corresponding to one of the major moons of the
planet. For Saturn's F ring as well as for the $\epsilon$ ring of
Uranus, perhaps a fifth body is also required to fully account for the
two shepherd moons.

\section{Conclusions and outlook}
\label{concl}

In this paper we have presented a general and self-consistent approach
for the occurrence of shepherded narrow planetary rings, and some
consequences for their structural properties. Our approach is based on
the local structure of phase space around stable (periodic or
quasi-periodic) solutions in regions where scattering dominates the
dynamics. The corresponding structure in phase space allows for a set
of positive measure to exhibit dynamical trapped motion, i.e. a
non-zero probability to find such trapped motion; particles with
initial conditions outside this small region escape to infinity
along scattering trajectories. The intrinsic rotation provides a
mechanism for dynamical confinement around the central planet. Here we
have considered that such rotation is induced by the perturbations of
a shepherd moon, though our approach is more general since the
intrinsic rotation can also be associated with interactions with
other, perhaps more massive, moons, or with the combined evolution of
the interacting many-body problem. A possible consequence is the
explanation of the non-shepherded narrow rings of Uranus: the
many-body problem is such that it defines {\it several} regions of
trapped motion, one for each specific narrow ring. An ensemble of
(non-interacting) particles will form a well-defined ring. Focusing in
regions where scattering dominates the dynamics allows to sharply
distinguish between trapped motion related to the occurrence of the
ring, and open motion manifested as regions with essentially no
material due to the escape of the particles.

Within our approach, the rings will be sharp-edged, as a consequence
of the scattering dynamics. The size of the stability region in the
extended phase space determines how narrow the rings are. For
comparatively small regions of trapped motion, the narrow ring will
typically be eccentric, since the particles will be close to the
organizing center, which in general displays such
eccentricity. Semi-analytical estimates of the edges of the ring can
be obtained by analyzing the bifurcation conditions that define the
region of bounded motion in the extended phase space (within the sea
of scattering trajectories) around the organizing center, at least for
models with two-degrees of freedom. Typically, the bifurcation
scenario that yields the regions of dynamically trapped motion
corresponds to the saddle--center bifurcation, and by consequence of
its fast development, the rings are narrow. We have illustrated our
considerations with three distinct examples, aiming to show explicitly
the robustness of the approach and its extension to and consequences
for a case with many degrees of freedom, and have proved that our
considerations hold also in the gravitational case as illustrated by
the RTBP without physical collisions. More realistic interactions,
e.g. including the oblateness of the planetary, can be easily
incorporated within our theory. While this does not yet explicitly
prove the connection to real narrow planetary rings, which are
complicated collisional many-body systems with gravitational and other
interactions, such an extension seems to us feasible, since our ideas
are based on general phase space considerations, and on results and
theorems which apply to Hamiltonian systems with more than two degrees
of freedom.

We emphasize that our construction of the rings is self-consistent, in
the sense that we do not introduce any ``cell'' or periodic boundary
conditions on a segment of the ring to make the problem amenable in
any sense. Therefore, radial and azimuthal structures do
appear. Moreover, the theoretical framework developed is not
restricted to short time scales, at least as long as the model
considered is believed to be applicable for longer time scales. It is
in this scope that our results may be relevant and contribute to the
understanding of real planetary narrow rings: the mechanism of
confinement is very efficient. We have explicitly showed which phase
space structures allow for the occurrence of such narrow
rings. Moreover, we have showed how rings develop further radial and
azimuthal structure: rings may display variable width and many
components which are entangled in a way that, at least qualitatively,
resembles the observed braids of Saturn's F rings. Within our
approach, the appearance of different strands and braids is not
related to the ring interparticle collisions, as
proposed~\citep{Hanninen1993}, nor to the existence of other
moonlets~\citep{LissauerPeale1986}; they are related to the
quasi-periodic character of the perturbing force of the rotating
interaction, as induced by the shepherd's explicitly eccentric
motion. This non-circular motion makes the Hamiltonian system
effectively have many degrees of freedom, which complicates its
analysis, but allows for other bifurcations as well as resonances. As
far as we are aware, this is a new explanation for the occurrence of
strands and braids, although it is somewhat in agreement with the idea
of associating periodic clumps with the eccentricity of shepherd
moons~\citep{ShowalterBurns1982}. The strands and braids that we obtain
display a periodic dynamical evolution, where the relevant time scale
is related to the time for repeated interactions, or close approaches,
with the shepherd. In our opinion, this points out the importance of
the quasi-periodic perturbation, i.e. the non-circular motion of the
intrinsic potential. This observation is in agreement with the results
of \citet{GiuliattiWinter00}, although these authors focus on
certain perturbations induced by the repeated interactions of a group
of particles with Prometheus within a cell of the ring. Whenever the
organizing periodic orbit closes after more than one full turn of the
secondary, the ring displays non-independent multiple components,
reflecting the shape of the organizing periodic or quasi-periodic
orbit in the rotating frame. Despite the fact that we have considered
only planar motion, and that this is again only a qualitative
comparison, the generalization of these ``higher periodic'' orbits to
a non-planar case serves as an interpretation of recent observational
findings~\citep{Charnoz05}. Moreover, preliminary
results~\citep{BMpreparation} show that our theory can also yield other
observed properties of narrow rings, such as arcs qualitatively
similar to those of Neptune~\citep{Namouni02}.

Our ultimate goal is to provide quantitative comparisons with observed
planetary rings. The results presented here are only qualitative due
to the fact that, so far, the examples are analogous to restricted
three-body like problems. Actually, the most realistic example we
considered is the circular (gravitational) RTBP. As we have discussed,
non-collision phase-space structures of the RTBP with extremely small
values of the mass parameter, which originate from the overwhelming
dominance of the central planetary attraction, make such simple
modeling definitely not realistic, since they yield two-body keplerian
wide circular rings. We have proposed to consider the interaction of
other planetary satellites and, perhaps too, other shepherds as a more
realistic model. The shepherds can be included in a first step as
zero-mass moonlets which break certain solutions of the many-body
problem (planet including oblateness, other moons and a particle of
the ring). Following the ideas of \citet{HitzlHenon77} for the RTBP
with $\mu=0$, periodic or quasi-periodic collision solutions with
respect to the shepherds can be generating orbits for the case of
small masses of the shepherds. Considering the stability regions in
the extended phase space when the shepherds have finite mass allows a
straightforward construction of the associated rings. Work
illustrating a many-body problem is in progress.

Our considerations may also be used to extend the applicability of
current models to longer time scales, once it has been showed that the
ring particles considered do belong to some regions of dynamically
bounded motion. Conversely, they may allow to qualify certain rings or
some of their structural properties as temporary phenomena, if a
realistic model does not have dynamically trapped particles. An
interesting question is whether chaotic motion of the shepherds may
still yield long-lived rings within our framework, that is, whether
they allow for bounded motion, and what the consequences are for their
structure. This is related to recent investigations where the motion
of Prometheus and Pandora seems to be
chaotic~\citep{Goldreich03a,Goldreich03b,Renner05}.

The strongest assumption within our approach is the non-interacting
character of the ring particles. This essentially allows us to
separate the many-body problem into a collection of independent
one-particle problems. This assumption must be taken as a first
approximation in the series expansion of the full many-body
Hamiltonian in terms of the mass of the ring particles. Here, the
particle--particle interactions among the ring particles will be
orders of magnitude smaller than other interactions, except maybe when
they collide. It is precisely the inelastic collisions among the ring
particles what we neglect. A possibility to include these collisions
in a simple way is with a random-walk model, taking place in all the
coordinates of the full phase space of the ring particles. Such a
model could also be used to describe dissipative effects like drag,
depending on the particular implementation. This random-walk approach
would certainly introduce an erosion mechanism of the ring, i.e., the
possibility of escape from the ring and a typical life-time; yet it
will preserve the properties that we discuss, such as the sharp-edges,
the eccentricity and the many components of the ring. This is because,
even when collisions are included, once a particle is outside the
region of trapped motion it escapes along a scattering
trajectory. Work in this direction is also in progress.

\begin{acknowledgements}
  We acknowledge financial support provided by the projects IN--101603
  (DGAPA--UNAM) and 43375 (CONACyT). We are thankful to \`Angel Jorba
  and Carles Sim\'o for discussions and comments, to Fr\'ed\'eric
  Masset and David Sanders for a critical reading and to the Referees
  for their suggestions. O.~Merlo is a postdoctoral fellow of the
  Swiss National Foundation (PBBS2--108932).
\end{acknowledgements}

\end{document}